\documentclass[10pt,journal]{IEEEtran}
\usepackage{amsmath,amssymb,amsfonts}
\usepackage{algorithmic}
\usepackage{array}
\usepackage[caption=false,font=normalsize,labelfont=sf,textfont=sf]{subfig}
\usepackage{textcomp}
\usepackage{stfloats}
\usepackage{url}
\usepackage{verbatim}
\usepackage{graphicx}
\usepackage{cite}
\def\BibTeX{{\rm B\kern-.05em{\sc i\kern-.025em b}\kern-.08em
    T\kern-.1667em\lower.7ex\hbox{E}\kern-.125emX}}
\usepackage{balance}
\usepackage[linesnumbered,ruled]{algorithm2e}
\usepackage{color, amsthm}
\usepackage{bm}
\usepackage{booktabs}
\usepackage{multirow}
\usepackage{makecell}
\usepackage{xcolor}
\usepackage{geometry}
\usepackage{setspace}

\newtheorem{corollary}{\bf Corollary}

\begin{document}

\title{Non-myopic Beam Scheduling for Multiple Smart Target Tracking in Phased Array Radar Network}

\author{Yuhang Hao, Zengfu Wang, Jos\'{e} Ni\~{n}o-Mora, Jing Fu, Min Yang, and Quan Pan
\thanks{Yuhang Hao is with the School of Automation, Northwestern Polytechnical University, and the Key Laboratory of Information Fusion Technology, Ministry of Education, Xi'an, Shaanxi, 710072, China. Part of this work is performed during his research stay in the Department of Statistics at Carlos III University of Madrid, Spain.
Zengfu Wang, Min Yang, Quan Pan are with the School of Automation, Northwestern Polytechnical University, and the Key Laboratory of Information Fusion Technology, Ministry of Education, Xi'an, Shaanxi, 710072, China.
Jos\'{e} Ni\~{n}o-Mora is with the Department of Statistics, Carlos III University of Madrid, Getafe, Madrid, 28903, Spain.
Jing Fu is with School of Engineering, RMIT University, Melbourne, VIC, 3000, Australia.
E-mail: (yuhanghao@mail.nwpu.edu.cn; wangzengfu@nwpu.edu.cn; jose.nino@uc3m.es; jing.fu@rmit.edu.au; sdta\_ym717@mail.nwpu.edu.cn; quanpan@nwpu.edu.cn).
(Corresponding author: Zengfu Wang.)
This work was in part supported by the National Natural Science Foundation of China~(grant no. U21B2008, 62233014). The work of J.\ Ni\~no-Mora was funded in part by Spain's State Research Agency (AEI) project PID2019-109196GB-I00/AEI/10.13039/501100011033.
}
}


\maketitle

\begin{abstract}
A smart target, also referred to as a reactive target, can take maneuvering motions to hinder radar tracking.
We address beam scheduling for tracking multiple smart targets in phased array radar networks.
We aim to mitigate the performance degradation in previous myopic tracking methods and enhance the system performance, which is measured by a discounted cost objective related to the tracking error covariance (TEC) of the targets.
The scheduling problem is formulated as a restless multi-armed bandit problem (RMABP) with state variables, following the Markov decision process.
In particular, the problem consists of parallel bandit processes.
Each bandit process is associated with a target and evolves with different transition rules for different actions, i.e., either the target is \emph{tracked} or not.
We propose a \emph{non-myopic}, scalable policy based on Whittle indices for selecting the targets to be tracked at each time.
The proposed policy has a linear computational complexity in the number of targets and the truncated time horizon in the index computation, and is hence applicable to large networks with a realistic number of targets.
We present numerical evidence that the model satisfies sufficient conditions for indexability (existence of the Whittle index) based upon partial conservation laws, and, through extensive simulations, we validate the effectiveness of the proposed policy in different scenarios.
\end{abstract}

\begin{IEEEkeywords}
Target tracking, beam scheduling, restless bandits, Whittle index.
\end{IEEEkeywords}

\section{Introduction}\label{sec:introduction}
\IEEEPARstart{S}{tate}-of-the-art sensor scheduling approaches have enabled higher flexibility for tracking smart targets through phased array radar networks, where real-time beam direction control is handled electronically \cite{haimovich2007mimo,zhang2022joint,yan2022radar}.
Dynamic tracking of smart targets, which have the ability to be aware that they are being tracked and adapt their dynamics accordingly to hinder tracking accuracy, has drawn recent research attention \cite{kreucher2006adaptive,savage2009sensor,zhang2015non}.
Hence, beam scheduling in radar networks with a trade-off between frequency of tracking and probability of maneuvering is complicated and of importance in smart-target tracking.
In this paper, we address a model to minimize the expected total discounted error for tracking smart targets through dynamic beam scheduling \cite{hero2007foundations,taner2012scheduling}.

Resource scheduling in radar systems has been mostly addressed through myopic policies.
In \cite{godrich2011sensor},
the challenge of antenna selection within a distributed multi-radar system was formulated as a knapsack problem, with the Cram\'{e}r--Rao lower bound (CRLB) being regarded as the objective function.
By exploring the multi-start knapsack tree, a local search algorithm with a multi-start strategy was adapted to solve the problem.
In \cite{zhang2022dynamic,zhang2020antenna,dai2022adaptive,yi2020resource,zhang2020joint,xie2017joint,shi2021joint,yan2020optimal,yan2020collaborative},
at each time, once obtaining the estimation of the target dynamic states,
the predicted one-step posterior CRLB (PCRLB) is calculated and used as the optimization metric to allocate resources for the next time step.
However, in general, myopic scheduling policies exhibit inevitable performance degradation in the long run~\cite{pang2019sensor}.

Non-myopic scheduling policies based on predicted multi-step objective functions or on \emph{partially observed Markov decision processes} (POMDPs) \cite{krishn16} have been investigated in a wide range of settings.
In \cite{pang2019sensor}, a bee colony algorithm with particle swarm optimization (PSO) was introduced to optimize a multi-step objective function representing operational risk.
In \cite{kreucher2004efficient,kreucher2006monte}, the joint multi-target probability density (JMPD) was recursively estimated by particle filtering methods, and the maximum expected R\'{e}nyi divergence between the JMPD and the JMPD recalculated with a new derived measurement was considered for long-term performance optimization.
In~\cite{kreucher2004efficient}, two information-directed approaches were presented to approximate the long-term effects of each action.
One is a path searching approach, aiming to reduce the computational complexity of a full Monte Carlo search.
The other one approximates an action's long-term benefit through a distinct function, expressed in relation to ``opportunity cost" or ``regret".
In \cite{kreucher2006monte}, a non-myopic sensor scheduling method leveraging the POMDP framework was utilized to improve long-term performance.
In \cite{gongguo2019non}, a smart target tracking problem was modeled as a POMDP using the PCRLB as system state and multi-step cost prediction over unscented sampling.
An improved decision tree search algorithm leveraging branch and bound was used to achieve non-myopic scheduling optimization for maneuvering target tracking.
Similarly as in \cite{gongguo2019non}, a branch-and-bound algorithm was considered in \cite{shan2020non} for sensor scheduling to reduce the interception probability.
In \cite{dong2021risk,shan2017non}, a POMDP-based branch-and-bound algorithm with worst-case exponential complexity was applied to a sensor allocation problem for reducing the risks of the sensor radiation interception and target threat level, which provided a non-myopic scheme with improved long-term performance.

Since computing optimal dynamic beam scheduling policies for POMDP multi-target tracking models is generally intractable and has worst-case exponential complexity, researchers have instead aimed to design suboptimal heuristic \emph{index} policies with low computational complexity.
An \emph{index policy} is based on defining an index for each target, a scalar function of its state, and giving higher tracking priority at each time to targets with larger index values.
A simple \emph{greedy index} for tracking a target moving in one dimension is its \emph{tracking error variance} (TEV), which is updated via scalar Kalman filter dynamics \cite{howard2004optimal}.

A versatile model for addressing optimal resource management problems and index policies are provided by the \emph{restless multi-armed bandit problem} (RMABP) \cite{whittle1988restless}, which is an extension to the \emph{multi-armed bandit problem} (MABP) \cite{gittins1979bandit,krishnamurthy2001hidden,gittins2011multi}.
The RMABP focuses on selecting a maximum of $M$ from a pool of $N \geq M$ \emph{projects} (or \emph{bandits}), which are versatile entities capable of being either \emph{active} (i.e., chosen) or \emph{passive} at each time.
Different from MABP, in an RMABP, a bandit is able to change state while passive.

While solving optimally the RMABP is generally intractable \cite{papTsik99}, Whittle proposed in \cite{whittle1988restless} a heuristic index policy, which has since been widely applied \cite{nmmath23}, where the index of a project depends only on its parameters.
The \emph{Whittle index} is implicitly defined as the state-dependent critical subsidy for passivity that makes both actions optimal in a single-project subproblem.
Yet, neither existence nor uniqueness of such a critical subsidy is guaranteed, so the Whittle index is well defined only for the class of \emph{indexable} projects.
Whittle conjectured that, when all projects are indexable, this index policy should approach optimality asymptotically under the average criterion as $M$ and $N$ grow to infinity in a fixed ratio.
\cite{weber1990index} showed this not to be generally true, but gave sufficient conditions under which the conjecture holds.

Establishing \emph{indexability} of a restless bandit model is widely considered a challenging task, which hinders the application of Whittle's index policy.
The currently dominant approach for proving indexability, mostly applied to one-dimensional state models, entails a two-step process: first, optimality of \emph{threshold policies} is proven for single-project subproblems; and, second, the Whittle index is obtained from the optimal threshold under an appropriate monotonicity condition on the latter.
Yet, some models have not yielded to such an approach, in particular restless bandit models for target tracking with Kalman filter dynamics, as considered here.

The RMABP formulation of multi-target tracking was first considered in \cite{la2006optimal}, extending work in \cite{howard2004optimal}.
Given the lack of effective tools to prove indexability in such a setting, \cite{la2006optimal} addressed the greedy index policy and focused on identifying sufficient conditions for its optimality in scenarios with two symmetric scalar-state targets.
\cite{dance2015kalman} further investigated that model using the prevailing approach to indexability, encountering that it posed seemingly unsurmountable difficulties, so they assumed the optimality of threshold policies without proving it.
The Whittle index policy has been studied via simulation to achieve near-optimality in related past studies \cite{liu2010indexability,nino2011sensor,wang2019whittle}.

Ni\~{n}o-Mora \cite{nino2001restless,nmmp02,nmmor06,nino2020verification} developed an alternative method to prove indexability and assess the Whittle index, circumventing the need to establish the optimality of threshold policies.
It sufficed to show that certain project performance metrics under such policies satisfy \emph{PCL-indexability conditions} ---named after satisfaction of \emph{partial conservation laws} (PCLs) \cite{nino2001restless,nmconsLawsEORMS}.
A \emph{verification theorem} guarantees the optimality of threshold policies as well as the indexability of the model and its Whittle index was given by an explicitly defined \emph{marginal productivity} (MP) \emph{index}.
This was proven in \cite{nino2001restless,nmmp02,nmmor06,nino2020verification} in increasingly general settings.

The PCL-indexability conditions for real-state restless bandits under the discounted criterion, which we shall apply here, were first outlined without proof in \cite{nino2008index}, and their application to multi-target tracking was numerically explored in \cite{nino2009multitarget,nino2016whittle}. The verification theorem was proven in \cite{nino2015verification,nino2020verification}.
\cite{dance2019optimal} used such conditions to first establish the indexability of scalar-state Kalman-filter restless bandits.

In the more practically relevant case of multi-target tracking RMABP models with multi-dimensional state Kalman filter dynamics, indexability is currently an open problem.
A heuristic approach was developed in \cite{yang2020restless}, where scalar-state project approximations were considered by taking as system state the trace of the channel estimation mean square error (MSE).

For smart targets, sensor scheduling problems become substantially more complex, as such targets react to radar sensing by switching their dynamics between different modes, e.g., a constant velocity (CV) mode, a constant acceleration (CA) mode, a constant turn (CT) mode, etc.~\cite{kreucher2006adaptive,savage2009sensor,zhang2015non,visina2018multiple}.
In such scenarios, it might be beneficial to refrain from sensing too often those targets that are likely to hide or escape when tracked, as this would make them harder to track in the future.
Therefore, proactive measures are necessary to effectively track smart targets.
In~\cite{kreucher2006adaptive}, when a smart target is viewed by a sensor, it reacts by ``hiding'' itself. The tracking problem was formulated as a Markov decision process (MDP) with an infinite time horizon, which was addressed through a two-stage reinforcement-learning approach with separated optimizations for detection and tracking.
In~\cite{savage2009sensor}, a modified algebraic Riccati equation (MARE) was applied in a Kalman-based target tracking system with smart targets.
This work considered game-theoretic methods that aimed to optimize waveform parameters and radar modes under imperfect measurement information, which was caused by the variation of the detection probability under target maneuvering.
In~\cite{zhang2015non}, a POMDP model with multiple dynamics models was considered to minimize the interception risk during the tracking of multiple reactive targets.
A roll-out method based on unscented transformation sampling (UTS) was introduced to approximate the long-term reward and to select sensors based on the closest distance policy.
In~\cite{nino2011sensor}, the problem of hunting hiding targets was formulated as an RMABP, where the state of a target is its posterior probability of being exposed.
Experimental results were presented showing that the Whittle index policy outperforms greedy policies.

Hence, substantial research challenges remain in sensor scheduling for tracking smart targets, including:
1) Solving optimally Markov decision process (MDP) models with multiple smart targets entails a prohibitively high computational cost;
2) In models with multi-dimensional \emph{tracking error covariance} (TEC) state, the application of the Whittle index policy is at present elusive; and
3) Increasing radar radiation for enhancing tracking performance needs to be optimally traded off with the potential of targets to react with evasive maneuvers in response to radiation detection.

In this paper, we consider a beam scheduling problem for tracking multiple smart targets that have a high probability of switching their dynamics when observed by phased array radars so as to hinder their tracking.
Otherwise, targets have a high probability of selecting a CV dynamics model under non-observation.
Similar to the interacting multiple models (IMM) method, the state of each target is defined as its sum-weighted TEC based on dynamics model probability vectors, which are assumed to be constant.
To formulate the beam scheduling problem, we consider an infinite-horizon discounted MDP model of RMABP type, where each target corresponds to a restless project.
The RMABP formulation is exploited to propose a non-myopic low-complexity scheduling policy.
In particular, we aim to apply the Whittle index policy to the dynamic beam scheduling for tracking smart targets. Yet, such a goal is hindered by the fact that
no proof of indexability (existence of the Whittle index) is at present available for this model.
We partially circumvent this difficulty by presenting some numerical evidence supporting the conjecture that the model satisfies the aforementioned PCL-indexability conditions, and use the MP index policy resulting from them. Recall that, if it were proven that the model satisfied such conditions, this would imply that the model is indexable and the MP index would be its Whittle index.
In addition, a different MP-based index policy is proposed in the multi-dimensional TEC state case.
Through extensive simulation results, this index policy is shown to outperform baseline policies.
The contributions of this paper are summarized as follows.
\begin{itemize}
\item
We formulate the beam scheduling problem for tracking multiple smart targets as an infinite-horizon discounted RMABP. The smart targets are characterized by tracking-action-dependent dynamics models with high maneuvering, which aim to hide themselves when being tracked. Each target is associated with a restless project, whose state is its sum-weighted TEC state based on the multiple-dynamic-model probability vectors.
\item
We aim to minimize the multi-target tracking error measured by the sum of the TECs. To mitigate the maneuvering of the smart targets, the system should avoid overuse of tracking action. In the real-state case, we apply the Whittle index policy (using the MP index in the belief that it coincides with the Whittle index
) to this beam scheduling problem.
\item The Whittle index policy is analyzed in the real-valued state model, while in the multi-dimensional TEC state case, a different MP index policy is considered. The TEC states evolve through the Kalman filter.
For the real-valued case, we present numerical evidence that the sufficient indexability conditions based on PCLs are satisfied, which is also used to evaluate the index. Furthermore, we provide the results of a simulation study assessing the sub-optimality gap of the Whittle index policy. The effectiveness of the proposed MP index policy for the multi-dimensional target state case is also assessed by a simulation study, where it is shown to outperform baseline policies.
\end{itemize}

The remainder of this paper is organized as follows.
In Section \ref{sec:model}, the target dynamics models and measurement models are defined.
Then, the state update under different actions and the discounted long-term objective of the scheduling problem are formulated.
In Section \ref{sec:solution}, the application of the Whittle index policy is discussed. The beam selection scheme is developed in the RMABP model based on   MP indices for real-valued and multi-dimensional state cases. The computational method of the lower bound of optimization functions and the computational complexity of policies are presented.
In Section \ref{sec:simulation}, real-valued and multi-dimensional state cases are considered, and reckless and cautious targets with different dynamics model probabilities are also considered. The simulation results demonstrate the superiority of the proposed index policies.
Section \ref{sec:conclusion} concludes the paper.

\section{Model description and problem formulation}\label{sec:model}
We update the target states via the Kalman filter and exploit the multiple dynamics model and corresponding model probabilities to represent the motion characteristics of smart targets.
Subsequently, we formulate the beam scheduling problem as an RMABP, of which each bandit process is associated with a target.

\subsection{Target dynamics models}\label{subsec:II-A}
\begin{figure}[!t]
\centering
\includegraphics[width=2.5in]{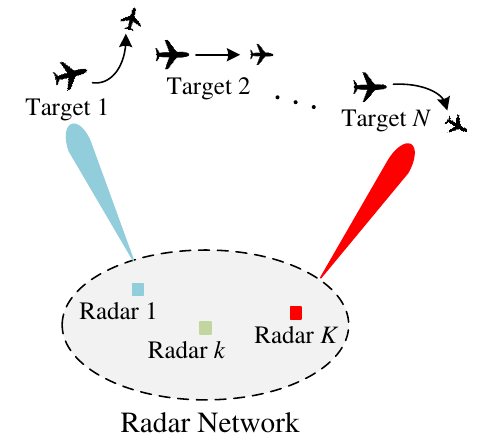}
\caption{System model of a phased array radar network.}
\label{figsce}
\end{figure}

We consider $N$ smart targets labeled by $n =1,2,\ldots, N $ and a radar network consisting of $K<N$ phased array radars, as illustrated in Fig.~\ref{figsce}.
In Fig.~\ref{figsce}, targets 1 and $N$ will likely change their dynamics models in response to radar tracking, while target 2 will be likely to maintain its dynamics model.
All radars are synchronized to operate over time slots $t=0,1,\ldots$.
The radar network centrally steers beams to track at most $K$ targets at each time, as one radar beam can be only assigned to one target in a time slot.

We assume that radars detect targets with unit probability.
Denote the kinematic state of target $n$ at time $t$ by $\bm{x}_{n,t}\in\mathbb{R}^L$.
Targets follow independent linear dynamics.
Meanwhile, they are smart, that is, they react to radar radiation by switching between $M$ dynamics models, labeled by $m = 1, 2, \ldots, M$, e.g., CV, CA, CT, etc.
Target $n$'s dynamics under model $m$ is
\begin{equation}\label{eq1}
\bm{x}_{n,t}=\bm{F}_{n}^m \bm{x}_{n,t-1}+\bm{w}_{n,t}^m.
\end{equation}
In \eqref{eq1}, $\bm{F}_{n}^m$ is the state transition matrix, and $\bm{w}_{n,t}^m$ is process noise with mean being zero and
covariance matrix being $\bm{\mathrm{Q}}_{n}^m$.
We assume that $\bm{\mathrm{Q}}_{n}^m=q_n^m\bm{\mathrm{Q}}_{n}$ for a covariance matrix $\bm{\mathrm{Q}}_{n}$, where
$0 < q_n^{1}<q_n^{2}<\cdots<q_n^{M}$ and $m=1$ corresponds to the CV model, so the uncertainty of the dynamics model increases with $m$.

At each time $t$, we use a binary variable $a_{n,t}\in\left\{0,1\right\}$ to denote the tracking action imposed on target
$n$.
Let $a_{n,t}=0$ if target $n$ is not tracked.
Conversely, when $a_{n,t}=1$, the target is actively tracked,
and a measurement $\bm{y}_{n,t}$ is acquired following the subsequent measurement equation,
\begin{equation}\label{eq3}
\bm{y}_{n,t}=\bm{\mathrm{H}}_n\bm{{x}}_{n,t} +\bm{v}_{n,t},
\end{equation}
where $\bm{\mathrm{H}}_n$ is the measurement matrix, and $\bm{v}_{n,t}$ is a zero-mean Gaussian white noise with covariance matrix $\bm{\mathrm{R}}_{n}$.

\subsection{Probabilities of dynamics models}\label{subsec:II-B}
Targets are assumed to change their dynamics models depending only on whether they are being tracked.
After action $a_n$ is applied on target $n$ at each time, its dynamics model is randomly drawn to be $m$ with probability $u_{n}^{a_n, m}$, for $m = 1, \ldots, M$.
We write $\bm{\mathrm{U}}_{n}\triangleq\left[\bm{u}_{n}^{0},\bm{u}_{n}^{1}\right]$, where $\bm{u}_{n}^{a_n}=\left[u_{n}^{a_n,1},\ldots,u_{n}^{a_n,M}\right]'$ and $(\cdot)'$ is the transpose operator.
We assume that $u_{n}^{0,1}>u_{n}^{0,m}$, $m=2,3,\ldots,M$, so targets that are not being tracked are more likely to move into model $m = 1$ (CV).
We further assume that $u_{n}^{1,1}<u_{n}^{1,m}$ for $m=2,3,\ldots,M$, so targets that are being tracked are less likely to move into model $m = 1$.

Consequently, the smart targets are characterized by the dynamics model probability matrix, and the states $\bm{x}$ of the targets are affected not only by the CV model but also by the other dynamics models.

\subsection{Kalman filter-based TEC update}\label{subsec:II-C}
Target TECs evolve based on the Kalman filter \cite{blair2023mse}.
Denote by ${\mathbf{P}}_{n, t}$ the TEC of target $n$ at the beginning of time slot $t$.
Starting from the initial TEC ${\mathbf{P}}_{n,0}$, ${\mathbf{P}}_{n, t}$ is recursively updated over time slots $t \geq 1$ being conditioned by the chosen actions, taking into account the dynamics models, as follows.

If the target is tracked at time $t-1$ ($a_{n,t-1}=1$), then
\begin{equation}\label{eq6}
\begin{aligned}
{\mathbf{P}}_{n,t}
=\mathrm{\phi}_{n}^{1}\left({\mathbf{P}}_{n,t-1}\right)
=\sum_{m=1}^M u_n^{1,m} \hat{\mathbf{P}}_{n,t}^{m},
\end{aligned}
\end{equation}
where
\begin{equation}\label{eq7}
\hat{\mathbf{P}}_{n,t}^{m}= \mathrm{\phi}_{n}^{1,m}\left({\mathbf{P}}_{n,t-1}\right)=\left(\bm{\mathrm{I}}-\bm{\mathrm{K}}_{n,t}^{m} \bm{\mathrm{H}}_n\right)\bar{\mathbf{P}}_{n,t|t-1}^m,
\end{equation}
\begin{equation}\label{eq8}
\bar{\mathbf{P}}_{n,t|t-1}^m=\bm{F}_{n}^m {\mathbf{P}}_{n,t-1}\left(\bm{F}_{n}^m\right)'+\bm{\mathrm{Q}}_{n}^m,
\end{equation}
\begin{equation}\label{eq9}
\bm{\mathrm{K}}_{n,t}^{m}=
\bar{\mathbf{P}}^m_{n,t|t-1}\bm{\mathrm{H}}'_n\left(\bm{\mathrm{H}}_n\bar{\mathbf{P}}_{n,t|t-1}^m\bm{\mathrm{H}}'_n+\bm{\mathrm{R}}_{n}\right)^{-1},
\end{equation}
and $\bm{\mathrm{I}}$ is the $L\times L$ identity matrix.

If the target is not tracked at time $t-1$ ($a_{n,t-1}=0$), then
\begin{equation}\label{eq10}
\begin{aligned}
{\mathbf{P}}_{n,t}
=\mathrm{\phi}_{n}^{0}\left({\mathbf{P}}_{n,t-1}\right)
=\sum_{m=1}^M u_n^{0,m} \bar{\mathbf{P}}_{n,t|t-1}^{m}.
\end{aligned}
\end{equation}

Note that the TEC update recursion given by (\ref{eq6})--(\ref{eq10}) is deterministic.
It follows that action $a_{n,t-1}$ updates the TEC ${\bm{\mathrm{P}}}_{n,t}$ through sum-weighted method based on the model probability vector $\bm{u}_{n}^{a_{n,t-1}}$.
When $a_{n,t-1}=1$, the target $n$ maneuvers with higher probability $u_{n}^{1,m}$, $m=2,3,\ldots,M$ than CV with lower probability $u_{n}^{1,1}$.
It leads to a larger determinant of TEC state ${\bm{\mathrm{P}}}_{n,t}$ than that of $\hat{\bm{\mathrm{P}}}_{n,t}^{1,1}$ solely calculated by CV model.
If $a_{n,t-1}=0$, then, since $q_n^1<q_n^m$, $m=2,3,\ldots,M$, with larger $u_{n}^{0,1}$ in \eqref{eq10}, the determinant of TEC state ${\bm{\mathrm{P}}}_{n,t}$ will be smaller than that with probability $\bm{u}_{n}^{1}$.

\subsection{RMABP model formulation}\label{subsec:II-D}
We next formulate the optimal beam scheduling problem for $K$ radars and $N$ smart targets as an infinite-horizon discrete-time RMABP, aiming to achieve the optimal dynamic selection of $K$ out of $N$ binary-action projects.
For such a purpose, we identify a target with a project, which can be operated under the active action (track the target) and the passive action (do not track it). We take as the state of project $n = 1, \ldots, N$ the corresponding target's TEC, $\mathbf{P}_{n,t}$, which moves over the state space $\mathbb{S}_{++}^{L}$ of symmetric positive definite $L \times L$ matrices, according to Kalman filter dynamics as shown in Section \ref{subsec:II-C}.

For target $n$ at time $t$, the immediate cost caused by imposing action $a_{n,t}$ is defined as
\begin{equation}\label{eq13}
C_n\left(\mathbf{P}_{n,t},a_{n,t}\right)\triangleq d_n\mathrm{tr}\left(\mathbf{P}_{n,t}\right)/L+h_{n} a_{n,t},
\end{equation}
where $\mathrm{tr}(\cdot)$ denotes the trace of a matrix, $d_n \ge 0$ is the target's weight, which is used to model the importance of targets, and $h_{n} \ge 0$ is a measurement cost.

The radar network selects at most $K$ out of $N$ smart targets to track in each time slot $t$.
We consider that each radar can only steer one beam to track one target at each time, which is formulated as
the constraint
\begin{equation}\label{eq12}
\sum_{n=1}^N a_{n,t} \leq K,~t=0,1,2,\ldots.
\end{equation}

Given the joint model probabilities $\bm{\mathbb{U}}=(\bm{\mathrm{U}}_n)_{n=1}^N$ and a TEC state at the beginning $\mathbf{\mathbb{P}}_0=(\mathbf{P}_{n,0})_{n=1}^N$,
we consider the dynamic optimization problem with discounted cost over an infinite time horizon.
\begin{equation}\label{eq14}
\begin{aligned}
\min_{\bm{\pi}\in\Pi(K)} &\mathrm{E}_{\mathbf{\mathbb{P}}_0}^{\bm{\pi}}\left[\sum_{t=0}^{\infty}\sum_{n=1}^{N}\beta^tC_n\left(\mathbf{P}_{n,t},a_{n,t}\right)\right],
\end{aligned}
\end{equation}
where $0<\beta<1$ is the discount factor, $\mathrm{E}_{\mathbf{\mathbb{P}}_0}^{\bm{\pi}}[\cdot]$ is expectation under policy $\bm{\mathrm{\pi}}$ conditioned on the initial state  $\mathbf{\mathbb{P}}_0$, and
denote by $\Pi(K)$ the set of stationary scheduling policies that satisfy \eqref{eq12}.
We denote by $V^*\left(\mathbf{\mathbb{P}}_0\right)$ the optimal cost of problem \eqref{eq14}.

\section{RMABP-based beam scheduling policy}\label{sec:solution}
Generally, determining an optimal policy for problem \eqref{eq14} is a computationally intractable task,
due both to the curse of dimensionality and to the continuous state space.
We next discuss how Whittle's \cite{whittle1988restless} approach for obtaining a heuristic index policy would apply to this smart target tracking problem
and discuss the challenges it poses. See also \cite{nino2009multitarget}.

\subsection{Problem relaxation and decomposition}\label{subsec:III-A}
To obtain a relaxed version of problem~\eqref{eq14},  we replace the sample-path constraint \eqref{eq12}
by the following constraint,
\begin{equation}\label{eq15}
\mathrm{E}_{\mathbf{\mathbb{P}}_0}^{\bm{\pi}}\left[\sum_{t=0}^{\infty}\sum_{n=1}^{N}\beta^t a_{n,t}\right]\leq \frac{K}{1-\beta}.
\end{equation}

Now, consistently with the above notation ${\Pi}(K)$,
let ${\Pi}(N)$ represent the set of stationary
scheduling policies capable of activating any quantity of projects (tracking any number of targets) at each time.
Then, a relaxation of problem~\eqref{eq14} is as follows.
\begin{equation}\label{eq16}
\begin{aligned}
\min_{\bm{\pi}\in{\Pi}(N)}~& \mathrm{E}_{\mathbf{\mathbb{P}}_0}^{\bm{\pi}}\left[\sum_{t=0}^{\infty}\sum_{n=1}^{N}\beta^tC_n\left(\mathbf{P}_{n,t},a_{n,t}\right)\right], \\
&\text{s.t.:}~\eqref{eq15}. \\
\end{aligned}
\end{equation}
Therefore, the minimum cost $V^{\textup{R}}(\mathbf{\mathbb{P}}_0)$ derived from problem \eqref{eq16} establishes a lower bound on $V^*(\mathbf{\mathbb{P}}_0)$.

Attach now a Lagrange multiplier $\lambda\geq 0$ to the aggregate constraint \eqref{eq15}.
The Lagrangian relaxation applied to \eqref{eq16} is
\begin{equation}\label{eq17}
\min_{\bm{\pi}\in{\Pi}(N)}\mathrm{E}_{\mathbf{\mathbb{P}}_0}^{\bm{\pi}}\left[\sum_{t=0}^{\infty}\sum_{n=1}^{N}\beta^t \{ C_n\left(\mathbf{P}_{n,t},a_{n,t}\right)+\lambda a_{n,t} \}
\right]-\frac{K\lambda}{1-\beta}
\end{equation}
Given a $\lambda \geq 0 $, the optimal value of the Lagrangian relaxation, denoted as $V^{\textup{L}}(\mathbf{\mathbb{P}}_0;\lambda)$, serves as a lower bound for $V^{\textup{R}}(\mathbf{\mathbb{P}}_0)$.

If we search for an optimal Lagrange multiplier $\lambda^*(\mathbf{\mathbb{P}}_0)$ based on the Lagrangian relaxation, the best lower bound for $V^{\textup{R}}(\mathbf{\mathbb{P}}_0)$ is obtained.
This defines the dual problem of \eqref{eq16}:
\begin{equation}\label{eq18}
V^{\textup{D}}(\mathbf{\mathbb{P}}_0)\triangleq \max_{\lambda\geq 0} \, V^{\textup{L}}\left(\mathbf{\mathbb{P}}_0;\lambda\right).
\end{equation}

Problem \eqref{eq17} can be decomposed into $N$ single-target independent subproblems given by
\begin{equation}\label{eq19}
\min_{\pi^n\in\Pi^n}\mathrm{E}_{\mathbf{P}_{n,0}}^{{\pi}^n}\left[\sum_{t=0}^{\infty}\beta^t \{ C_n\left(\mathbf{P}_{n,t},a_{n,t}\right)+\lambda a_{n,t} \}
\right].
\end{equation}
In \eqref{eq19}, $\Pi^n$ is the set of stationary scheduling policies corresponding to target $n$.
Here, multiplier $\lambda$ has the economic interpretation of an additional tracking cost.

\subsection{Computing the Lower bound $V^{\textup{D}}(\mathbf{\mathbb{P}}_0)$}\label{subsec:III-D}
In the one-dimensional case with real-valued state space for a single bandit process, $V^{\textup{D}}\left((P_{n,0})_{n=1}^N\right)$ can be calculated through the value iteration approach \cite{brown2020index}.
In general, consider the lower bound $V^{\textup{D}}(\mathbf{\mathbb{P}}_0)$ in \eqref{eq18} which is the optimal value of the Lagrangian dual problem. Recall that we aim to minimize the expected total discounted (ETD) cost. Based on the Bellman equation, we firstly derive the optimal value function for the process $\{\mathbf{P}_{n,t},t=0,1,\ldots\}$ of target $n$ in \eqref{eq19}. Given the Lagrange multiplier $\lambda$, we obtain
\begin{equation}\label{eq31}
\begin{aligned}
v^{\lambda,*}_{n}\left(\mathbf{P}_{n,t}\right)
=\min\limits_{a_{n,t}\in\{0,1\}}
\Biggl\{ & C_n\left(\mathbf{P}_{n,t},a_{n,t}\right)+\lambda a_{n,t} \\
+&\beta\mathrm{E}\left[v^{\lambda,*}_{n}\left(\phi_1^{a_{n,t}}\left(\mathbf{P}_{n,t}\right)\right) \right]\Biggl\}.
\end{aligned}
\end{equation}

Subsequently given the Lagrange multiplier $\lambda$ in \eqref{eq17}, we can derive the multitarget optimal value function $L^{\lambda,*}\left(\mathbf{\mathbb{P}}_{t}\right)$ and substitute \eqref{eq31} into it, as given by
\begin{equation}\label{eq32}
\begin{aligned}
&L^{\lambda,*}\left(\mathbf{\mathbb{P}}_{t}\right) \\
=&\min\limits_{\bm{a}_{t}\in\{0,1\}^N}\left\{\sum\limits_{n=1}^N C_n\left(\mathbf{P}_{n,t},a_{n,t}\right)+\beta \mathrm{E}\left[L^{\lambda,*}\left(\tilde{\phi}^{\bm{a}_{t}}\left(\mathbf{\mathbb{P}}_{t}\right) \right) \right] \right.\\
&\left.+\lambda\beta\left(\sum_{n=1}^{N}\left(a_{n,t}-K\right)\right)\right\}\\
=&\sum_{n=1}^{N}v^{\lambda,*}_{n}\left(\mathbf{P}_{n,t}\right)-\frac{K\lambda}{1-\beta},
\end{aligned}
\end{equation}
where
\begin{equation}\label{eq32add}
\tilde{\phi}^{\bm{a}_t}\left(\mathbf{\mathbb{P}}_{t}\right)
=\Bigl(\phi_1^{a_{1,t}}\left(\mathbf{P}_{1,t}\right), \ldots, \phi_N^{a_{N,t}} \left(\mathbf{P}_{N,t}\right)\Bigr),
\end{equation}
representing the next-slot joint target states.

Based on the established value function and a given $\lambda$, we establish the $k$-th iteration of value function $v^{\lambda}_{n,k}\left(\mathbf{P}_{n,t}\right)$ as
\begin{equation}\label{eq33}
\begin{aligned}
v^{\lambda}_{n,k}\left(\mathbf{P}_{n,t}\right)
=\min\limits_{a_{n,t}\in\{0,1\}}&\Biggl\{ C_n\left(\mathbf{P}_{n,t},a_{n,t}\right)+ \lambda a_{n,t}  \\
&+\beta\mathrm{E}\left[\mathbb{I}_{a_{n,t}=1} v^{\lambda}_{n,k-1}\left(\phi^{1}_n (\mathbf{P}_{n,t})\right)\right] \\
&+\beta\mathrm{E}\left[\mathbb{I}_{a_{n,t}=0} v^{\lambda}_{n,k-1}\left(\phi^{0}_n (\mathbf{P}_{n,t})\right)\right] \Biggl\}.
\end{aligned}
\end{equation}

Then, we use the value iteration approach to continue the iterative process over the state space until convergence of values on diverse joint state $\mathbf{\mathbb{P}}_{t}$. Finally, we obtain the optimal value $v^{\lambda,*}_{n}\left(\mathbf{P}_{n,t}\right)$ in \eqref{eq31} and derive $L^{\lambda,*}\left(\mathbf{\mathbb{P}}_{t}\right)$ in \eqref{eq32}.
Given the initial joint state $\mathbf{\mathbb{P}}_0$ and a given $\lambda$, the minimum cost $V^{\textup{L}}(\mathbf{\mathbb{P}}_0;\lambda)$ can be computed by $L^{\lambda,*}\left(\mathbf{\mathbb{P}}_{0}\right)$.
The lower bound $V^{\textup{D}}\left(\mathbf{\mathbb{P}}_0\right)$ of the dual problem in \eqref{eq18} can be computed by searching a $\lambda\in(0,\infty)$ to satisfy
\begin{equation}\label{eq34}
V^{\textup{D}}(\mathbf{\mathbb{P}}_0)\triangleq \max_{\lambda\geq 0} \, L^{\lambda,*}\left(\mathbf{\mathbb{P}}_{0}\right).
\end{equation}

Above all, we regard the lower bound of problem \eqref{eq18} as the optimal cost of the original problem \eqref{eq14} to further calculate the suboptimality gap of policies and numerically analyze the near optimality of the Whittle index policy, given a constant ratio $\xi=K/N$.

\subsection{Indexability and Whittle index policy}\label{subsec:III-B}
Consider now the following structural property of subproblems \eqref{eq19}, referred to as \emph{indexability} in \cite{whittle1988restless}, which we formulate next as in \cite[Section 3.2]{nino2016whittle}.
Note that target $n$'s subproblem \eqref{eq19} is \emph{indexable} if there exists an \emph{index}
$\lambda_n^*\colon \mathbb{S}_{++}^L \to \mathbb{R}$
that characterizes its optimal policies, as follows:
for any $\lambda\in\mathbb{R}$, when the target is in state $\mathbf{P}_{n}$,
$\lambda_n^*(\mathbf{P}_{n}) >\lambda$~($\lambda_n^*(\mathbf{P}_{n}) \leq\lambda$) is necessary and sufficient conditions for that taking action
$a_{n}=1$~($a_{n}=0$) is optimal.

If each subproblem were indexable, the \emph{Whittle index policy} would be to
track up to $K$ out of the $N$ targets with larger index values, among those with nonnegative indices.
If the current index of a target is negative, it is not tracked.

Yet, at present it is unknown whether restless projects with multi-dimensional  Kalman filter dynamics such as those above are indexable, even for a single dynamics model ($M = 1$).
Indexability has only been established (in \cite{dance2019optimal}) for the special case of target tracking with
scalar Kalman filter dynamics and a single dynamics model, by applying the PCL-indexability approach for
real-state projects in \cite{nino2020verification}.

\subsection{PCL-indexability approach for real-state targets}\label{subsec:III-C1}
We next outline the PCL-indexability approach for real-state restless bandit projects as it applies to the present model (so $L = 1$).
In the sequel we pay attention to the MP index of a single target with multiple dynamics models as described above, and drop the subscript $n$ from notations.

Starting from an initial state $P_0=P \in \mathbb{S}_{++} \triangleq (0,\infty)$, for a policy $\pi$,
define the \emph{cost metric}
\begin{equation}\label{eq20_1}
F(P, \pi)\triangleq\mathrm{E}^{\pi}_{P} \left[\sum_{t=0}^{\infty}\beta^tC\left(P_t,a_t\right)\right],
\end{equation}
which gives the ETD cost, where $P$ is the target's \emph{tracking error variance} (TEV) state.

The \emph{work metric}  is defined as
\begin{equation}\label{eq21_1}
G(P, \pi)\triangleq\mathrm{E}^{\pi}_{P}\left[\sum_{t=0}^{\infty}\beta^t a_t\right].
\end{equation}

The subproblem \eqref{eq19} can be reformulated as
\begin{equation}\label{eq22_1}
\min_{\pi\in\Pi} \, F(P, \pi)+\lambda G(P, \pi).
\end{equation}

The indexability of subproblem \eqref{eq22_1} is studied under deterministic stationary policies and threshold policies.
Deterministic stationary policies are represented as (Borel measurable) subsets of TEV states,
where the corresponding target is tracked.
We will employ an \emph{active set} denoted as $B \subseteq \mathbb{S}_{++}$ to characterize the \emph{$B$-active policy}.
In particular, for any given threshold level $z \in \bar{\mathbb{S}}\triangleq \mathbb{S}_{{++}}\cup\{-\infty,\infty\}$, we will designate this as the
\emph{$z$-threshold policy}.
This means that, for a target in TEV state $P$,
if $P>z$, the target is tracked; otherwise,
it is not tracked.
Therefore, a $z$-threshold policy has active set $B(z) \triangleq \{P \in \mathbb{S}_{++} : P > z\}$.
It is clear that, if $0< z< \infty$, then
$B(z) = (z,\infty)$;
if $z\leq0$, then $B(z) = \mathbb{S}_{++}$;
and if $z=\infty$, then $B(z) = \emptyset$.
We shall denote the corresponding cost and work metrics in \eqref{eq20_1} and \eqref{eq21_1} by $F(P,z)$ and $G(P,z)$, respectively.

Given a threshold $z$, the cost and work metrics are characterized by the functional equations.
\begin{equation}\label{eq23_1}
F(P, z)=
\begin{cases}
C\left(P,1\right)+\beta F\left(\phi^{1}(P),z\right), & P > z \\
C\left(P,0\right)+\beta F\left(\phi^{0}(P),z\right), & P \leq z.\\
\end{cases}
\end{equation}
and
\begin{equation}\label{eq24_1}
G(P, z)=
\begin{cases}
1+\beta G\left(\phi^{1}(P),z\right), & P > z \\
\beta G\left(\phi^{0}(P),z\right), & P\leq z.\\
\end{cases}
\end{equation}
In practice, such metrics can be approximately computed by a value-iteration scheme with a finite \emph{truncated time horizon} $\tau$, as discussed in \cite[Section 11]{nino2020verification}.


Next, we represent as $\langle a,z\rangle$ the policy that executes action $a$ at time $t = 0$ and subsequently follows the $z$-threshold policy from $t = 1$ onwards.
Accordingly, we introduce the corresponding marginal metrics for the previously discussed measures.
Specifically, we define the
\emph{marginal cost metric} $f(P, z)\triangleq F(P, \langle 0, z\rangle)-F(P, \langle 1, z\rangle)$ and the
\emph{marginal work metric} $g(P, z)\triangleq G(P, \langle 1, z\rangle)-G(P, \langle 0, z\rangle)$.
If $g(P, z) > 0$, we further define the MP metric function by
\begin{equation}\label{eq25_1}
\mathrm{mp}(P, z)=\frac{f(P, z)}{g(P, z)}.
\end{equation}

If $g(P, P) > 0$ for all $P$, we define the \emph{MP index} by
\begin{equation}\label{eq26_1}
\mathrm{mp}^*(P)\triangleq\mathrm{mp}\left(P,P\right).
\end{equation}

As in \cite[Definition 7]{nino2020verification}, we say that the subproblem \eqref{eq22_1} is \emph{PCL-indexable} (with respect to threshold policies) if the following
\emph{PCL-indexability} conditions hold:
\begin{itemize}
\def\labelenumi{\arabic{enumi})}
\item (PCLI1) $g(P, z)>0$ for every state $P$ and threshold $z$;
\item (PCLI2)
$\mathrm{mp}^*(P)$ is monotone non-decreasing, continuous, and bounded below;
\item (PCLI3) for each $P$, the metrics $F(P, z)$, $G(P, z)$ and the index $\mathrm{mp}^*(P)$ are related by: for $-\infty < z_1 < z_2 < \infty$,
\[
F(P, z_2) - F(P, z_1) = \int_{(z_1, z_2]} \mathrm{mp}^*(z) \, G(P, dz),
\]
where the right-hand side is a Lebesgue--Stieltjes integral.
\end{itemize}

The interest of the above PCL-indexability conditions lies in their applicability through the \emph{verification theorem} in \cite[Theorem 1]{nino2020verification}, which ensures that, for a
real-state project, conditions (PCLI1)--(PCLI3) above imply that the project is indexable, and the MP index $\mathrm{mp}^*(P)$ is its Whittle index.
\begin{corollary}\label{C1}
If subproblem \eqref{eq22_1} is PCL-indexable, then it is indexable with Whittle index $\mathrm{mp}^*(P)$.
\end{corollary}

This was the approach deployed in \cite{dance2019optimal} to prove indexability for the special case of a single dynamics model ($M = 1$).
Yet, proving that conditions (PCLI1)--(PCLI3) hold for the present model with $M \geq 2$ is beyond the scope of this work.
Still, we conjecture that the model satisfies such conditions, and will present partial numerical evidence supporting satisfaction of such conditions.
We shall further use the MP index above as a surrogate of the Whittle index, and refer to it as the Whittle index. Note that only if it were proven that conditions (PCLI1)--(PCLI3) hold could we ensure that the MP index is indeed the Whittle index.

\subsection{A heuristic MP index for multi-dimensional state targets}\label{subsec:III-C2}
In the case of target dynamic states moving over in a $L$-dimensional state space $\mathbb{R}^L$, with $L \geq 2$, the TEC matrix indicates the tracking accuracy of the target state's components, e.g., location and velocity. Since the TEC state is an $L \times L$ matrix, a convenient scalar measure of tracking performance is given by its trace.
The application of the PCL-indexability approach for proving indexability and evaluating the Whittle index in such a case remains an open problem, even in the case of single target dynamics ($M = 1$).
This section leverages the past success of the PCL-indexability and extends a new heuristic approach to define an MP index for the multi-dimensional case.

In particular, we define the cost metric $F(\mathbf{P},\pi)$ and the work metric $G(\mathbf{P},\pi)$ by replacing the one-dimensional $P$ in \eqref{eq20_1} and \eqref{eq21_1} with the multi-dimensional $\mathbf{P}$. The target’s optimal tracking subproblem \eqref{eq19} becomes
\begin{equation}\label{eq22_2}
\min_{\pi\in\Pi} \, F\left(\mathbf{P},\pi\right)+\lambda G\left(\mathbf{P},\pi\right).
\end{equation}

We need to define the meaning of threshold policies in the present setting.
For such a purpose, we shall consider that the $z$-threshold policy tracks the target in state $\mathbf{P}$ if and only if $\mathrm{tr}(\mathbf{P})/{L} > z$.
Then, the corresponding cost metric $F(\mathbf{P}, z)$ and work metric $G(\mathbf{P},z)$
are the unique solutions to the following functional equations:
\begin{equation}\label{eq23_2}
F\left(\mathbf{P}, z\right)=
\begin{cases}
C(\mathbf{P}, 1)+\beta F(\phi^{1}(\mathbf{P}), z), &
\mathrm{tr}(\mathbf{P})/{L} > z \\
C(\mathbf{P},0)+\beta F(\phi^{0}(\mathbf{P}), z), &
\mathrm{tr}(\mathbf{P})/{L} \leq z.\\
\end{cases}
\end{equation}
\begin{equation}\label{eq24_2}
G(\mathbf{P}, z)=
\begin{cases}
1+\beta G(\phi^{1}(\mathbf{P}), z), & \mathrm{tr}(\mathbf{P})/L > z \\
\beta G(\phi^{0}(\mathbf{P}, z)), & \mathrm{tr}(\mathbf{P})/L\leq z.\\
\end{cases}
\end{equation}
Again, we can approximate such solutions through a value-iteration approach using a truncated time horizon $\tau$.

Now, similar to the scalar case above, we define the marginal cost metric $f(\mathbf{P}, z)$, the marginal work metric $g(\mathbf{P}, z)$, and the MP metric
\begin{equation}\label{eq25_2}
\mathrm{mp}(\mathbf{P}, z)=\frac{f(\mathbf{P}, z)}{g(\mathbf{P}, z)}=\frac{F(\mathbf{P}, \langle 0, z\rangle)-F(\mathbf{P}, \langle 1,z\rangle)}{G(\mathbf{P},\langle 1,z\rangle)-G(\mathbf{P},\langle 0,z\rangle)},
\end{equation}
provided that $g(\mathbf{P}, z) > 0$.

The corresponding MP index is defined by
\begin{equation}\label{eq26_2}
\mathrm{mp}^*(\mathbf{P})\triangleq \mathrm{mp}(\mathbf{P},\mathrm{tr}(\mathbf{P})/L),
\end{equation}
provided that $g(\mathbf{P}, \mathrm{tr}(\mathbf{P})/L) > 0$ for all $\mathbf{P}$.

However, there is currently no theory supporting any relation of the index $\mathrm{mp}^*(\mathbf{P})$ with the Whittle index in the multi-dimensional case, assuming that both are well defined.
Still, we will discuss in Section~\ref{sec:simulation} the results of a simulation study on the performance of the index policy based on the heuristic MP index $\mathrm{mp}^*(\mathbf{P})$.

\subsection{Scheduling scheme based on the Whittle index policy}\label{subsec:III-E}
In Sections~\ref{subsec:III-C1} and \ref{subsec:III-C2}, we discussed the MP-based indices for a bandit process with real-valued and multi-dimensional state spaces, respectively.
In general, the MP-based index policy prioritizes tracking specific targets based on their state-dependent MP indices, regardless of the dimensions of the state, in descending order.
The pseudo-code of implementing the index policy with given indices is provided in Algorithm \ref{algorithm:Algorithm 1}, which is applicable to both cases with one-dimensional and multi-dimensional state spaces.
\begin{algorithm}[!t]
\DontPrintSemicolon
\SetAlgoLined
\caption{The scheduling scheme based on the Whittle index policy}
\label{algorithm:Algorithm 1}
\tcp*[l]{\textbf{Initialization}}
The model parameters of the problem, i.e., $N$, $K$, $T$, $\beta$, $\bm{\mathbb{Q}}_{n}\triangleq\left(\bm{\mathrm{Q}}_{n}\right)^N_{n=1}$, $\bm{\mathbb{R}}_{n}\triangleq\left(\bm{\mathrm{R}}_{n}\right)^N_{n=1}$, and the target parameters, i.e. $\mathbf{\mathbb{P}}_{0}$, $\bm{\mathbb{U}}$. \\
\tcp*[l]{\textbf{Main loop}}
\For{$t \gets 1$ \KwTo $T$}{
\For{$n \gets 1$ \KwTo $N$}{
Calculate the index of the TEC state $\mathbf{P}_{n,t}$ of arm $n$ in \eqref{eq26_2};\;
}
Select the $K$ targets with the largest indexes and generate the $a_{n,t}$, $n=1,\ldots,N$, where for the multi-target with the same indexes, they are randomly selected;\;
\For{$n \gets 1$ \KwTo $N$}{
Update the TEC state $\mathbf{P}_{n,t+1}$ of arm $n$ in \eqref{eq6} and \eqref{eq10};\;
}
Calculate the cost in time $t$.
}
\end{algorithm}

\subsection{Computational complexity analysis} \label{subsec:III-F}
For the case with real-valued bandit states, the complexity for computing the indices in \eqref{eq26_2} is only linear in the number of targets $N$ and the truncated time $\tau$.
Nonetheless, for the case with multi-dimensional state space for each bandit process, the linearity does not hold in general due to the complex matrix operations in \eqref{eq6} and \eqref{eq10}.

In the Whittle index policy, the truncated time horizon $\tau$ is the recursive horizon in~\eqref{eq23_2} and \eqref{eq24_2}.
If the actions of each target across $\tau$ time horizon are all assumed as 0, the computational complexity of \eqref{eq25_2} is $2N\tau(n_{\phi^0}+4)+2+1$. Otherwise, the computational complexity is $2N\tau(n_{\phi^1}+4)+2+1$, where $n_{\phi^1}$ and $n_{\phi^0}$ ($n_{\phi^0}< n_{\phi^1}$) represent the complexity of updating matrix states when the corresponding action variable is set to $1$ and $0$, respectively.
Due to the diverse combination of actions across $\tau$ time horizon, the computational complexity of the Whittle index is less than $2N\tau(n_{\phi^1}+4)+2+1$. That is, we have a complexity of $O(N\tau)$.

We introduce two greedy approaches as baseline policies.
One is the \emph{TEC index} policy, which defines the indices by
\begin{equation}\label{index_tec}
\lambda^{\text{TEC}}(\mathbf{P})=d \, \mathrm{tr}(\mathbf{P})/L.
\end{equation}
The computational complexity of the TEC index is $O(N)$.
The other prioritizes actions according to the MP indices with $\beta=0$, namely, the \emph{myopic index} policy with the indices given by
\begin{equation}\label{index_my}
\lambda^{\text{myopic}}=d \left[\mathrm{tr}(\phi^0(\mathbf{P}))-\mathrm{tr}(\phi^{1}(\mathbf{P}))\right]/L.
\end{equation}
The computational complexity for $\mathrm{mp}^{\text{myopic}}$ is
$O(N(n_{\phi^1}+n_{\phi^0}))$.
The baseline policies exhibit lower computational complexity but, in general, fail to take consideration of long future effects of the employed actions.
Based on numerical results in \cite{nino2009multitarget,nino2016whittle,nino2011sensor}, with appropriate $\beta$ ($\beta\leq 0.9$), the MP indices $\mathrm{mp}^*(\bm{P})$ (described in \eqref{eq26_2}) with truncated horizon $\tau=100$ is sufficiently close to the analytical Whittle indices, for which the computational complexity is comparable to the greedy approaches.

\section{Simulation and Analysis}\label{sec:simulation}
In this section, we report on numerical experiments to assess the satisfaction of the PCL-indexability conditions and the effectiveness of the proposed Whittle (MP) index policy in the real-state case. We will compare the proposed index policy with the two greedy baselines, i.e., the TEC policy and the myopic policy, in the scalar and the multi-dimensional state cases.

\subsection{PCL-indexability conditions and MP index evaluation for real-state smart targets}\label{subsec:IV-A}
We start by considering the PCL-indexability conditions stated in Section \ref{subsec:III-C1} as they apply to scalar state targets. Recall that they
are sufficient conditions for indexability (existence of the Whittle index) and further give an explicit expression of the Whittle index in the form of an MP index.
Establishing analytically that the present model satisfies such conditions is beyond the scope of this work.
Still, we present below a sample of numerical evidence supporting the conjecture that the model satisfies such conditions.

To test the conditions numerically, we consider targets with $M = 2$ dynamics models,
which we refer to as CV ($m = 1$) and CT ($m = 2$), and parameter values
$F^{\text{CV}}=1.1$, $F^{\text{CT}}=1.3$, $Q^{\text{CV}}=1$, $Q^{\text{CT}}>Q^{\text{CV}}$, $H=1$, and $R=2$.
There are no measurement costs ($h=0$) and the weight is $d=1$.
We assume two types of smart targets, called \emph{reckless} and \emph{cautious}, where reckless targets have a higher probability of maneuvering (using the CT model) when tracked than cautious targets.
Hence, the dynamics model probability matrix $\bm{\mathrm{U}}=\left[\bm{u}^{0},\bm{u}^{1}\right]$ differs with the target types. Specifically, we take
\begin{equation}\label{eqU}
\bm{\mathrm{U}}=\begin{cases}
\left[[0.90, 0.10]',[0.20, 0.80]'\right], & \text{for reckless targets}\\
\left[[0.95, 0.05]',[0.60, 0.40]'\right], & \text{for cautious targets.}\\
\end{cases}
\end{equation}
where, if target $n_1$ belongs to reckless targets, target $n_2$ belongs to cautious targets, we can obtain $u_{n_1}^{0,m}>u_{n_2}^{0,m}$ and $u_{n_1}^{1,m}>u_{n_2}^{1,m}$, $m=2,\ldots,M$.

In the following numerical experiments, the infinite series defining the metrics of interest have been approximately computed by using the truncated time horizon $\tau=100$.
The discount factor is $\beta=0.9$, and the TEV state $P$ has been taken from a grid of values, each separated by a distance of $10^{-2}$.

\begin{figure}[!t]
    \begin{minipage}[t]{0.5\linewidth}
        \centering
        \includegraphics[width=\textwidth]{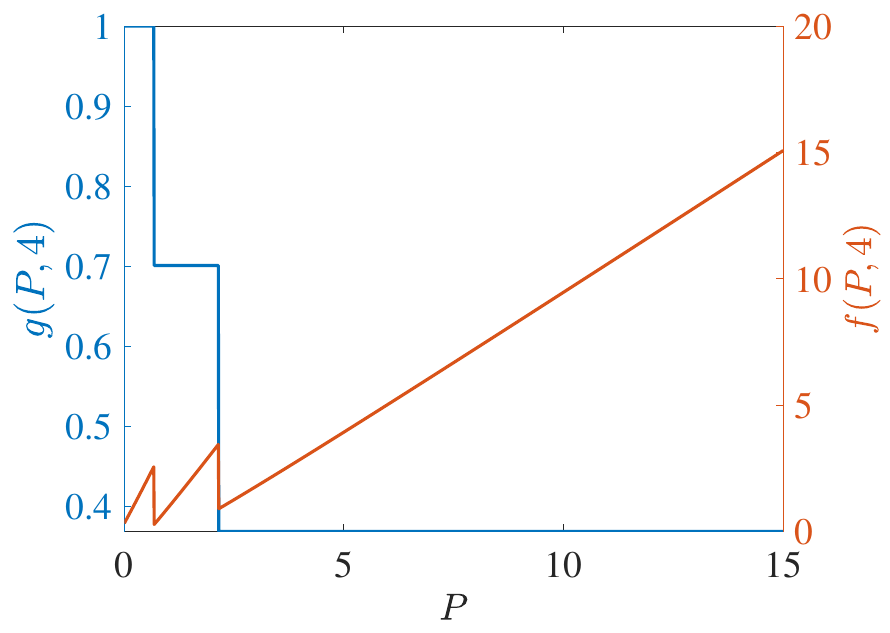}
        \centerline{(a)}
    \end{minipage}%
    \begin{minipage}[t]{0.5\linewidth}
        \centering
        \includegraphics[width=\textwidth]{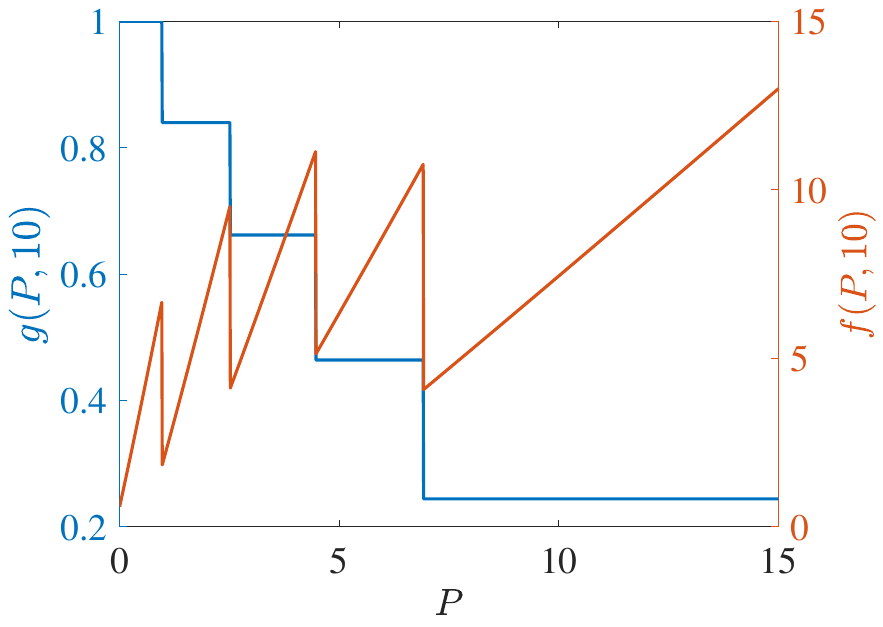}
        \centerline{(b)}
    \end{minipage}

    \begin{minipage}[t]{0.5\linewidth}
        \centering
        \includegraphics[width=\textwidth]{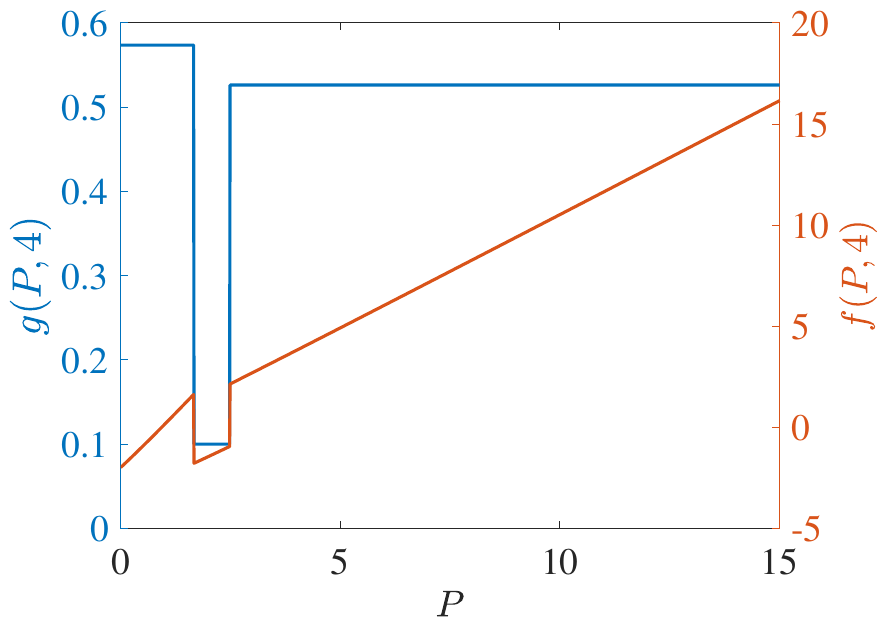}
        \centerline{(c)}
    \end{minipage}%
    \begin{minipage}[t]{0.5\linewidth}
        \centering
        \includegraphics[width=\textwidth]{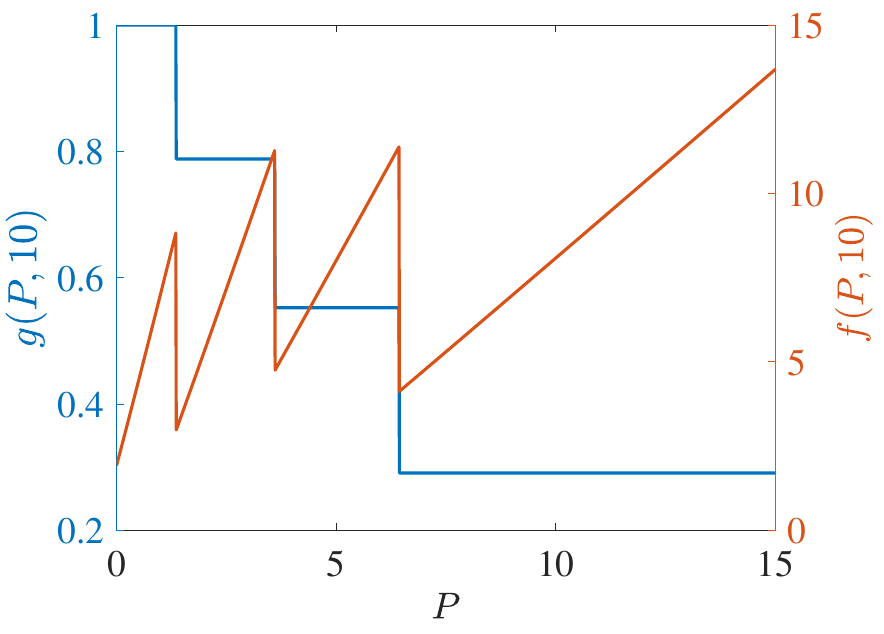}
        \centerline{(d)}
    \end{minipage}%
    \caption{Marginal work metric $g(P, z)$ and marginal cost metric $f(P, z)$ for different $z$ and $Q^{\text{CT}}$ for reckless targets: (a) $z=4$ and $Q^{\text{CT}}=4$; (b) $z=10$ and $Q^{\text{CT}}=4$; (c) $z=4$ and $Q^{\text{CT}}=10$; (d) $z=10$ and $Q^{\text{CT}}=10$.}
    \label{fig1-rec}
\end{figure}

\begin{figure}[!t]
    \begin{minipage}[t]{0.5\linewidth}
        \centering
        \includegraphics[width=\textwidth]{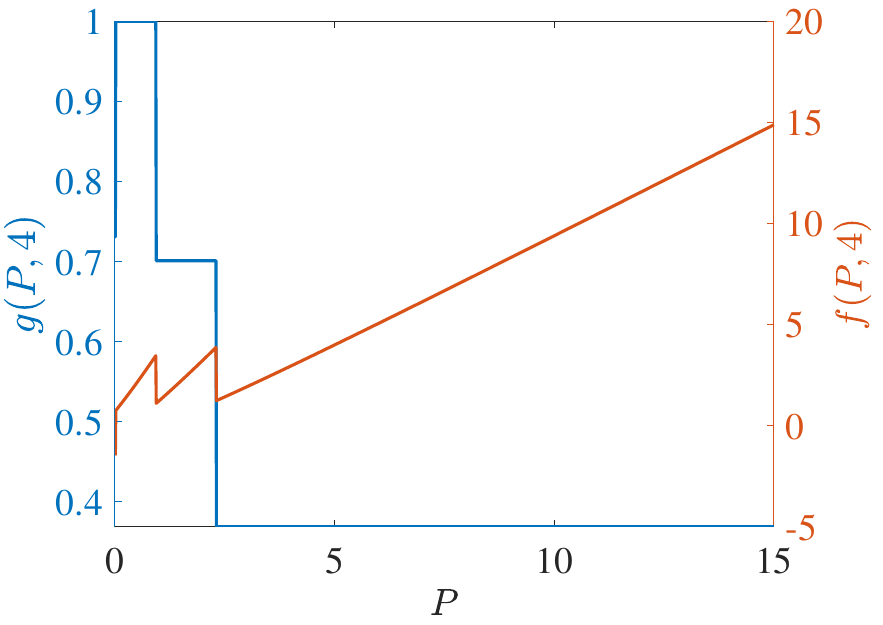}
        \centerline{(a)}
    \end{minipage}%
    \begin{minipage}[t]{0.5\linewidth}
        \centering
        \includegraphics[width=\textwidth]{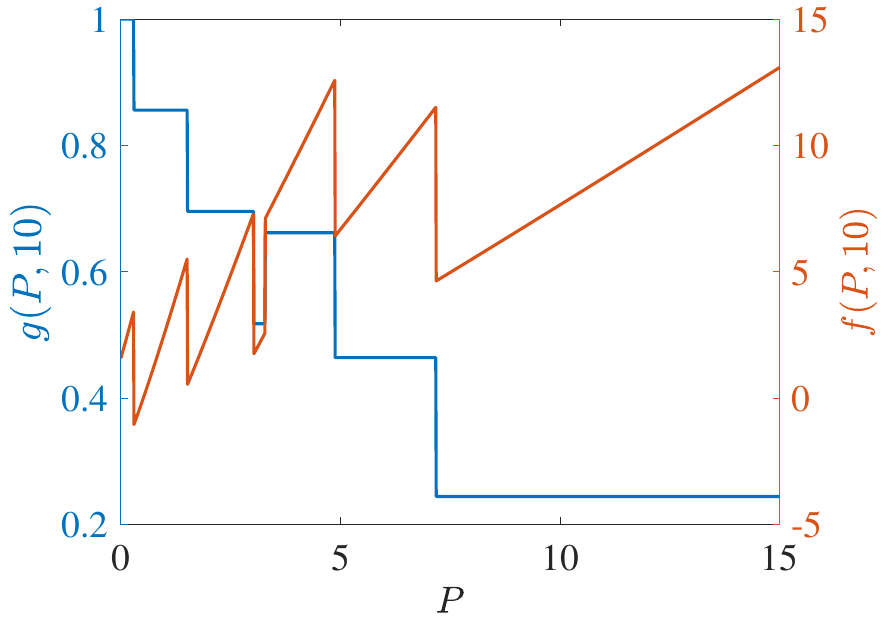}
        \centerline{(b)}
    \end{minipage}

    \begin{minipage}[t]{0.5\linewidth}
        \centering
        \includegraphics[width=\textwidth]{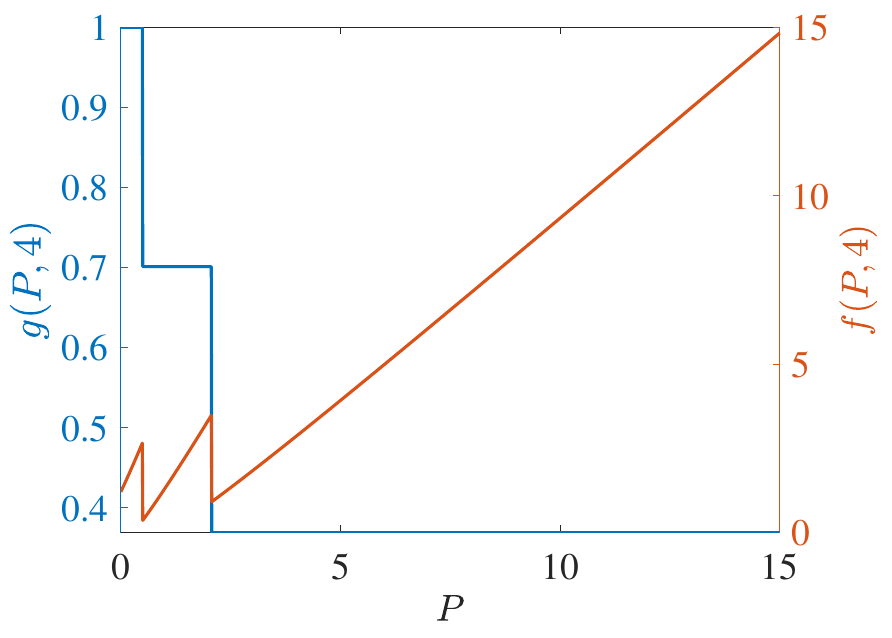}
        \centerline{(c)}
    \end{minipage}%
    \begin{minipage}[t]{0.5\linewidth}
        \centering
        \includegraphics[width=\textwidth]{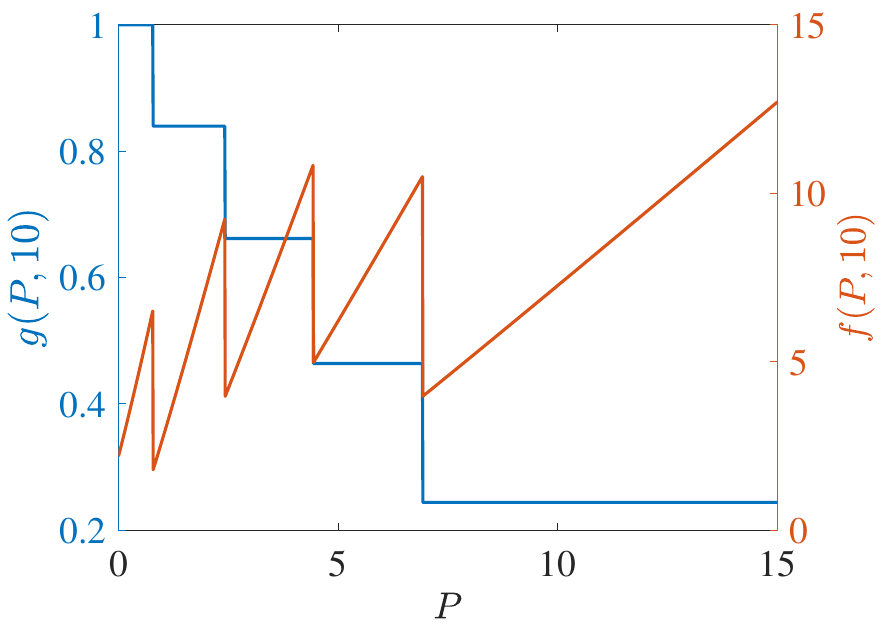}
        \centerline{(d)}
    \end{minipage}%
    \caption{Marginal work metric $g(P, z)$ and marginal cost metric $f(P, z)$ for different $z$ and $Q^{\text{CT}}$ for cautious targets: (a) $z=4$ and $Q^{\text{CT}}=4$; (b) $z=10$ and $Q^{\text{CT}}=4$; (c) $z=4$ and $Q^{\text{CT}}=10$; (d) $z=10$ and $Q^{\text{CT}}=10$.}
    \label{fig1-cau}
\end{figure}

Start with PCL-indexability condition (PCLI1), namely that marginal work metric $g(P, z)$ be positive for each TEV state $P$ and threshold $z$.
Figs.\ \ref{fig1-rec} and \ref{fig1-cau} plot both $g(P,z)$ and the marginal cost metric $f(P,z)$ against $P$ for several choices of threshold $z$ and of the $Q^{\text{CT}}$ parameter, for reckless targets and cautious targets in Fig.~\ref{fig1-rec} and Fig.~\ref{fig1-cau}, respectively.
The plots support the validity of condition (PCLI1), as $g(P, z) > 0$ in each instance, a result that has been observed under other parameter choices not reported here.

\begin{figure}[!t]
    \begin{minipage}[t]{0.5\linewidth}
        \centering
        \includegraphics[width=1.75in]{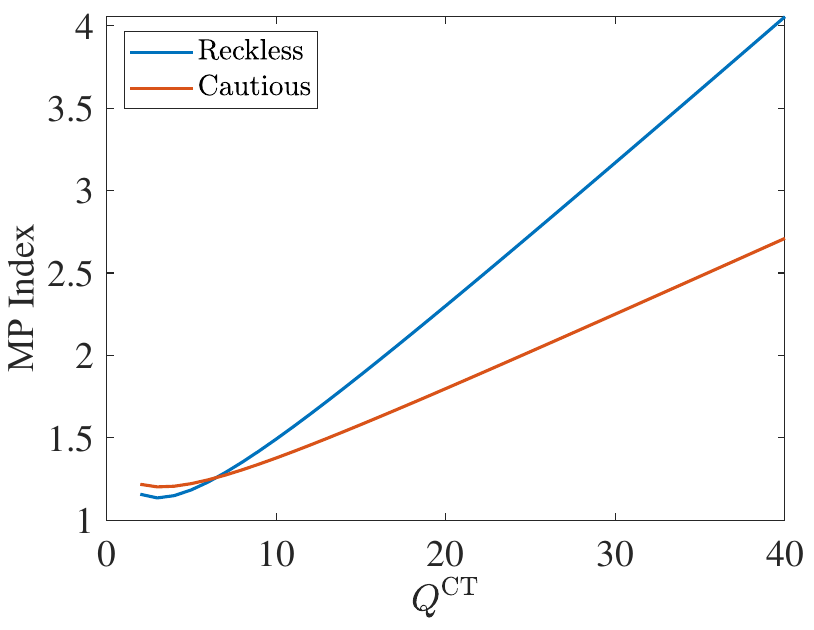}
        \centerline{(a)}
    \end{minipage}%
    \begin{minipage}[t]{0.5\linewidth}
        \centering
        \includegraphics[width=1.73in]{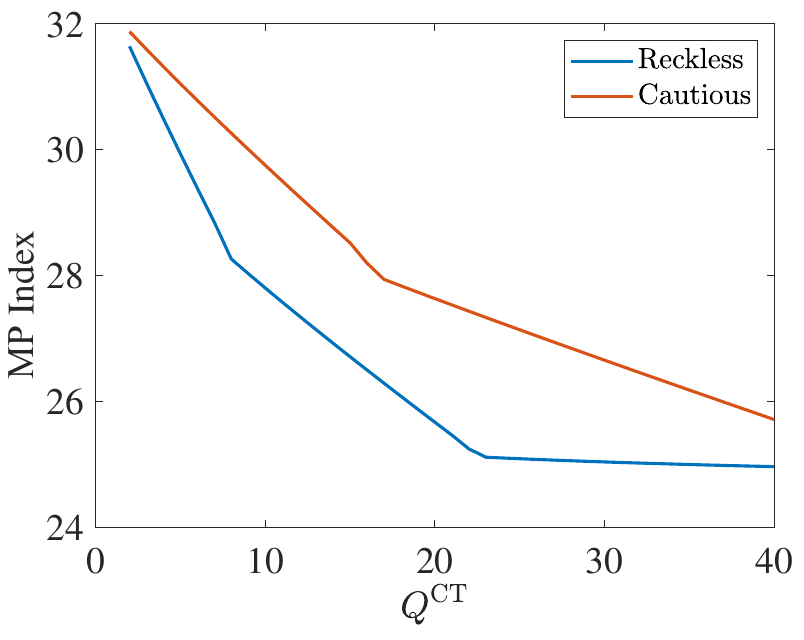}
        \centerline{(b)}
    \end{minipage}
    \begin{minipage}[t]{0.5\linewidth}
        \centering
        \includegraphics[width=1.75in]{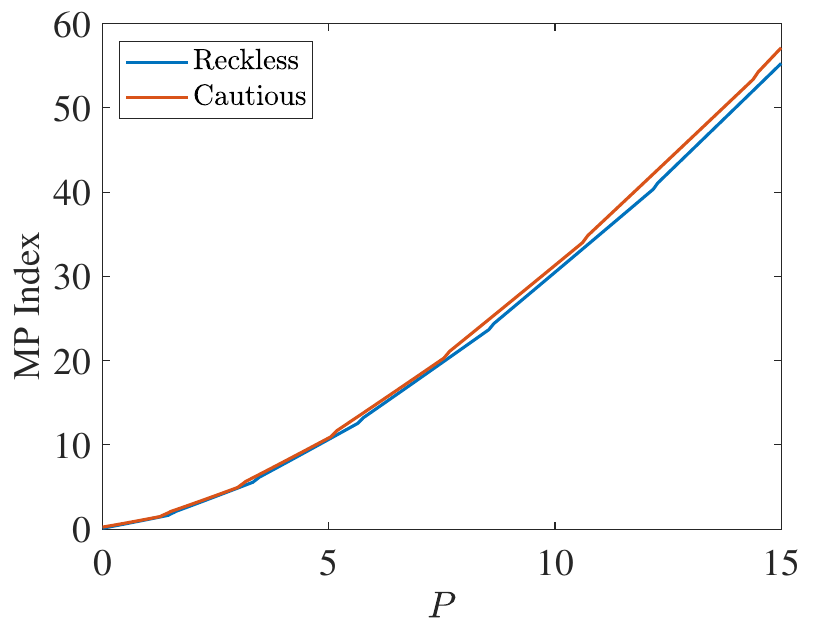}
        \centerline{(c)}
    \end{minipage}%
    \begin{minipage}[t]{0.5\linewidth}
        \centering
        \includegraphics[width=1.75in]{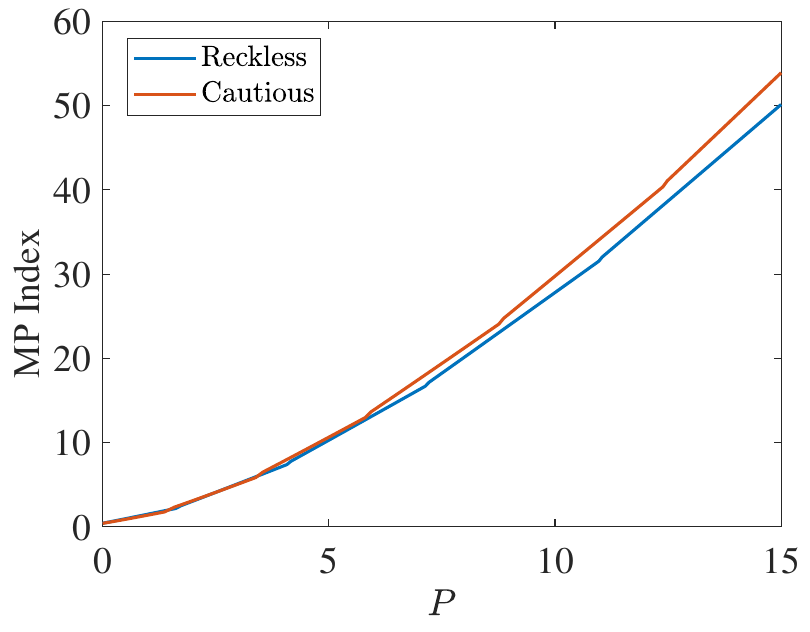}
        \centerline{(d)}
    \end{minipage}%
    \caption{MP index for different target types with different process noise variances $Q^{\text{CT}}$ and states $P$: (a) $P=1$; (b) $P=10$; (c) $Q^{\text{CT}}=4$; (d) $Q^{\text{CT}}=10$.}
    \label{fig2}
\end{figure}

Regarding PCL-indexability condition (PCLI2), namely that the MP index defined by $\mathrm{mp}^*(P) \triangleq f(P, P)/g(P, P)$ be bounded below, continuous and monotone non-decreasing,
Fig.~\ref{fig2} plots the MP index with choices of $Q^{\text{CT}}$ and $P$ for the two target types considered.
Figs.~\ref{fig2} (c) and (d) show the MP index vs.\ $P$ for $Q^{\text{CT}}=4,10$, respectively. The plots are clearly consistent with condition (PCLI2), which has been observed in other instances not reported here.

Fig.~\ref{fig2} (a) and (b) further show the MP index vs.\ $Q^{\text{CT}}\in[0,40]$ for states $P=1,10$, respectively.
When the state $P$ is small, the long-term cost is predominantly influenced by $Q^{\text{CT}}$. Consequently, with the same $P=1$, the reckless target with a higher $Q^{\text{CT}}$ will obtain a higher MP index value.
Otherwise, with a large $P=10$, prioritizing the long-term cost minimization, the MP index policy will sacrifice the immediate cost at the current time and track the target with higher $Q^{\text{CT}}$ after a certain number of time steps.
That is, the target with a lower $Q^{\text{CT}}$ is tracked currently. Hence, the MP index will decrease over $Q^{\text{CT}}$.
Due to the higher probability of a cautious target acquiring a lower cost with the same state $P$ and $Q^{\text{CT}}$ under action 1, cautious targets obtain a higher index than reckless targets in all states, which indicates that the MP index policy takes the tracking action prudently for reckless targets.

In summary, the above partial numerical evidence that the model of concern satisfies PCL-indexability conditions (PCLI1) and (PCLI2), and hence we shall further use the MP index as a surrogate of the Whittle index, and refer to it as the Whittle index.

\subsection{Performance results with real-valued states} \label{subsec:IV-B}
In this subsection, $N=8$ smart targets are tracked by the radars network with $K=1,\ldots,3$ radars. For each target, the radar tracking time horizon $T=100$ s and $\beta=0.9$. We set up the Whittle index policy with the truncated time $\tau=100$, the TEV index policy where $\lambda^{\text{TEV}}(P)=d*P$
, and the myopic policy that degenerates to $\mathrm{mp}^{\text{myopic}}=d*(\phi^0(P, \bm{u}^0)-\phi^{1}(P, \bm{u}^1))$ to assess tracking performances.

The parameters $F_n^{\text{CV}}$, $F_n^{\text{CT}}$, $Q_n^{\text{CV}}$, $h_n$ of target $n$ and $H$, $R$ are assumed the same as Section \ref{subsec:IV-A}, $n=1,\ldots,N$.
The initial state $P_{n,0}\sim\mathrm{U}(0,2)$.
Then, we define three scenarios with only reckless type, cautious type, and each type of target accounting for 4, where the process noise variance $Q_n^{\text{CT}}=\{2,2,\ldots,2\}$ and $\{2,3,\ldots,9\}$ in the first two scenarios, respectively; Otherwise, $Q_n^{\text{CT}}=\{2,2,\ldots,2\}$ and $\{2,3,4,5,2,3,4,5\}$ in the third scenario.
We assume $d_n=5$, $n=1$ and $d_n=1$, $n\neq1$ in the first two scenarios. Moreover, $d_n=5$ for reckless targets and $d_n=1$ for cautious targets in the third scenario.

Table~\ref{tab:q1_rec}, \ref{tab:q1_cau}, \ref{tab:q1_rec_cau} show the discounted objective performance in three policies.
Each policy is analyzed by varying the optional number $K$ of radars and $N_{mc}=100$ Monte Carlo simulations.
When all the targets are homogeneous and heterogeneous, the Whittle index policy can obtain the best scheduling performance and outperform the TEV index and myopic policies.
The myopic policy considers the one-step cost optimization, which is beneficial to the different types of targets; While the TEV index policy is not insensitive to the variation of targets.
Consequently, the performance of the myopic policy is better than the TEV policy.
Hence, the simulation results of all the scenarios show the superiority of the Whittle index policies.
\begin{table}[!t]
\begin{center}
\caption{Results with different numbers of radars for all the same reckless targets}
\label{tab:q1_rec}
\begin{tabular}{ c | c | c | c | c }
\hline
$Q^{\text{CT}}_n$ & Policies & $K=1$ & $K=2$ & $K=3$ \\
\hline
\multirow{3}{*}{\makecell[c]{$\{2,2,\ldots,2 \}$}}
& Whittle index & 823.19 & 400.53 & 284.65 \\
& myopic & 868.71 & 405.85 & 293.80 \\
& TEV & 871.19 & 406.17 & 293.72 \\
\hline
\multirow{3}{*}{\makecell[c]{$\{2,3,\ldots,9 \}$}}
& Whittle index & 961.25 & 458.49 & 319.84\\
& myopic & 993.93 & 464.43 & 326.04 \\
& TEV & 1009.15 & 465.12 & 334.06 \\
\hline
\end{tabular}
\end{center}
\end{table}

\begin{table}[!t]
\begin{center}
\caption{Results with different numbers of radars for all the same cautious targets}
\label{tab:q1_cau}
\begin{tabular}{ c | c | c | c | c }
\hline
$Q^{\text{CT}}_n$ & Policies & $K=1$ & $K=2$ & $K=3$ \\
\hline
\multirow{3}{*}{\makecell[c]{$\{2,2,\ldots,2 \}$}}
& Whittle index & 750.91 & 377.36 & 268.30 \\
& myopic & 790.61 & 381.92 & 275.81 \\
& TEV & 790.40 & 384.06 & 275.75 \\
\hline
\multirow{3}{*}{\makecell[c]{$\{2,3,\ldots,9\}$}}
& Whittle index & 817.23 & 406.26 & 285.67 \\
& myopic & 849.91 & 409.88 & 296.35 \\
& TEV & 861.88 & 410.56 & 296.19 \\
\hline
\end{tabular}
\end{center}
\end{table}

\begin{table}[!t]
\begin{center}
\caption{Results with different numbers of radars for different type targets}
\label{tab:q1_rec_cau}
\begin{tabular}{ c | c | c | c | c }
\hline
$Q^{\text{CT}}_n$ & Policies & $K=1$ & $K=2$ & $K=3$ \\
\hline
\multirow{3}{*}{\makecell[c]{$\{2,2,\ldots,2 \}$}}
& Whittle index & 1554.35 & 772.19 & 547.38 \\
& myopic & 1614.41 & 807.35 & 567.14 \\
& TEV & 1622.97 & 808.89 & 567.94 \\
\hline
\multirow{3}{*}{\makecell[c]{$\{2,3,4,5,$ \\ $2,3,4,5\}$}}
& Whittle index & 1664.83 & 821.05 & 581.47 \\
& myopic & 1731.69 & 859.02 & 605.75 \\
& TEV & 1733.95 & 860.42 & 605.73 \\
\hline
\end{tabular}
\end{center}
\end{table}

\subsection{Near optimality with real-valued states}\label{subsec:IV-C}
Within this section, we validate the near optimality of the Whittle index policy discussed in Section \ref{subsec:III-B} through different $d_n$ and $Q_{n}^{\text{CT}}$,
when the number of radars and targets increases with a fixed ratio $\xi=K/N=1/4$.
The parameters $F_n^{\text{CV}}$, $F_n^{\text{CT}}$, $Q_n^{\text{CV}}$, $h_n$ of target $n$ and $H$, $R$ are assumed the same as Section \ref{subsec:IV-B}.

We first assume three scenarios, where $d_n=1$, $n=1,\ldots,N$.
Three scenarios are divided by the types of targets, including all the reckless targets, all the cautious targets, and each type of target accounts for $50\%$.
For each target $n$, let ${{P}}_{n,0}=0.01$.
In addition, the process noise variances of the three scenarios are $Q_{n}^{\text{CT}}=\{2,\ldots, N+1\}$, $Q_{n}^{\text{CT}}=\{2,\ldots, N+1\}$, $Q_{n}^{\text{CT}}=\{2,\ldots, N/2+1,2,\ldots, N/2+1\}$, $n=1,\ldots, N$, respectively.
We modify the simulation population by letting $K$ vary.
For each scenario, the Whittle index truncates the corresponding infinite series to $\tau=100$ terms with $\beta=0.9$, and performances are evaluated by $N_{mc}=100$ Monte Carlo simulations.

\begin{figure*}[!t]
\centering
    \begin{minipage}[t]{0.32\linewidth}
        \centering
        \includegraphics[width=2.2in]{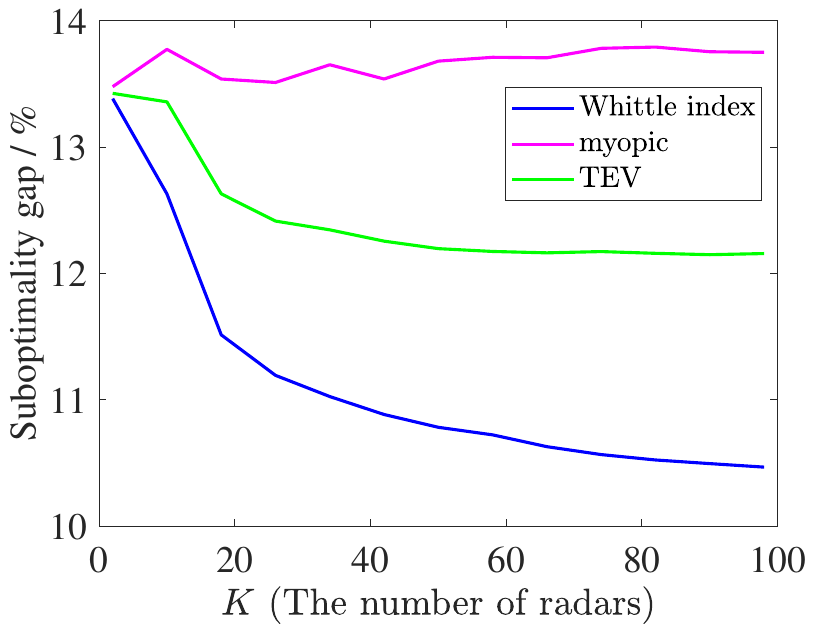}
        \centerline{(a)}
    \end{minipage}%
    \begin{minipage}[t]{0.32\linewidth}
        \centering
        \includegraphics[width=2.24in]{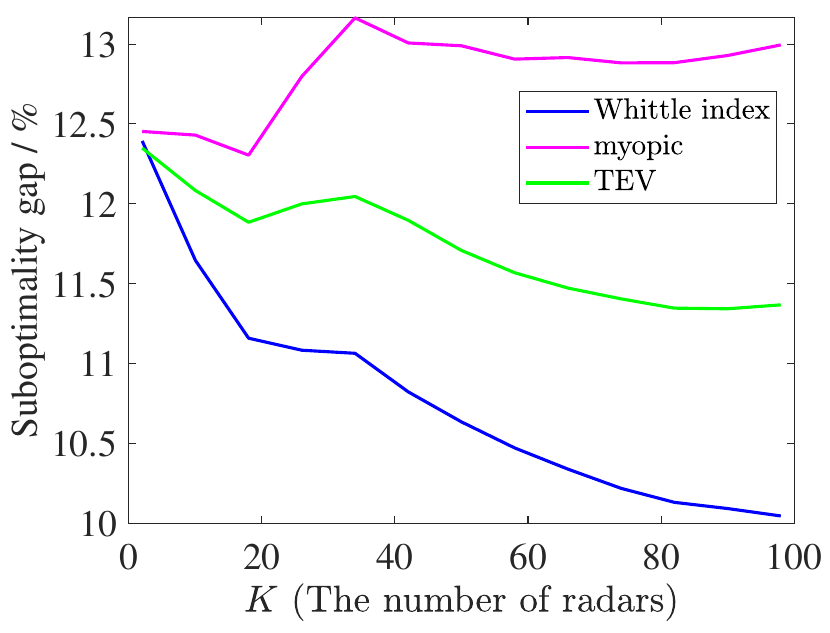}
        \centerline{(b)}
    \end{minipage}%
    \begin{minipage}[t]{0.32\linewidth}
        \centering
        \includegraphics[width=2.2in]{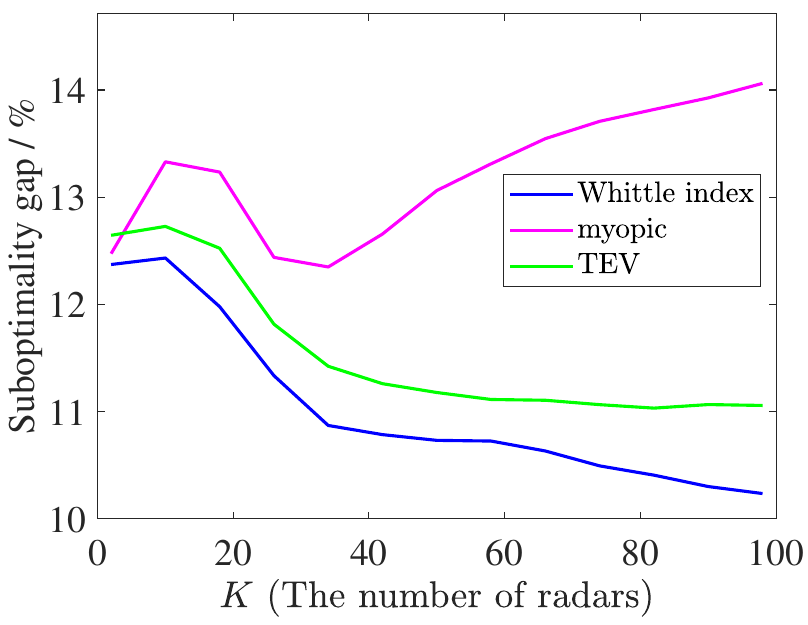}
        \centerline{(c)}
    \end{minipage}%
    \caption{The suboptimality gap of policies for heterogeneous targets. (a) Reckless targets. (b) Cautious targets. (c) Two types of targets.}
    \label{fig3}
\end{figure*}

Through the calculations of the lower bound in Section~\ref{subsec:III-D}, we derive the simulation results about the suboptimality gap between the lower bound per target and the objective per target of index policies, as shown in Fig.~\ref{fig3}, respectively.

\begin{figure}[!t]
\centering
\includegraphics[width=2.2in]{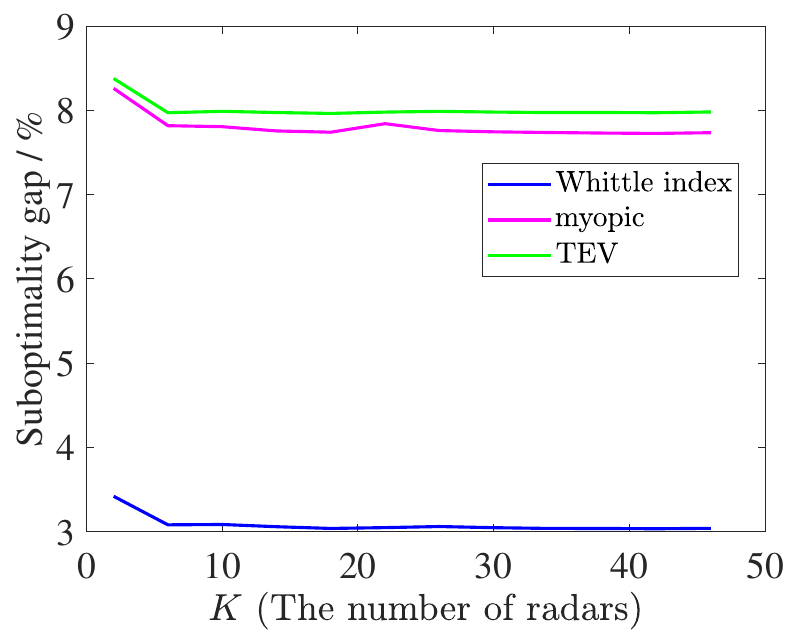}
\caption{The suboptimality gap in the constant ratio of Two types of targets.}
\label{fig5}
\end{figure}

Different from the above three scenarios, we assume another scenario with the ratio $\xi=K/N=1/4$. Firstly, there are $50\%$ reckless targets with $d_n=5$ in the population of $N$ and others are cautious with $d_n=1$. Secondly, we assume that all of the targets share the identical process noise variance $Q_{n}^{\text{CT}}=4$.
Other simulation parameters of each target are the same as the above three scenarios. The initial state for each target $n$ is assumed to be ${{P}}_{n,0}\sim\mathrm{U}(0,2)$. We modify the simulation population by letting $K$ vary. The results are shown in Fig.~\ref{fig5}.

Fig.~\ref{fig3} shows that the suboptimality gaps of the Whittle index policy and the TEV index policy decrease as the radar system scale increases. However, the suboptimality gap of the Whittle index policy converges to a lower level, which is lower than $10.5\%$, while the TEV index policy obtains a larger suboptimality gap. Unfortunately, the myopic policy has the worst performance and the suboptimality gap may increase as $K$ grows.
In Fig.~\ref{fig5}, the Whittle index policy also achieves the lowest suboptimality gap, which is stable at about $3.0\%$. While the TEV index policy and the myopic policy obtain $8.0\%$ and $7.8\%$, respectively.
Consequently, the Whittle index policy obtains lower objectives per target than other policies and the lowest suboptimality gap in all four scenarios.

Above all, the near optimality of the Whittle index policy is validated in this beam scheduling problem.
When all the targets are homogeneous in different groups related to the target type and target parameters in Fig.~\ref{fig5}, except ${{P}}_{n,0}\sim\mathrm{U}(0,2)$, and the suboptimality gap of the Whittle index policy is approximately stable at the lowest level.
Accordingly, the Whittle index policy can efficiently solve the beam scheduling problem with real-valued TEV states and achieve better non-myopic performance than other greedy policies.

\subsection{Performance results with multi-dimensional states}
In this subsection, we attempt to expand the application of the MP index policy to the multi-dimensional states case.
$N=8$ targets are tracked by the radar network with $K=1,\ldots,3$ radars. For each target, the radar tracking time horizon $T=100$ s.
The MP index policy, the TEC index policy, and the myopic policy are described in Section \ref{subsec:III-F} to assess the tracking performance.

Let $\bm{x}_{n,t}=\left[x_{n,t}, \dot{x}_{n,t}, y_{n,t}, \dot{y}_{n,t}\right]'$ be the dynamic state of target $n$ at time $t$, where $[x_{n,t}, y_{n,t}]$ represents the position of target $n$ at time $t$, $[\dot{x}_{n,t}, \dot{y}_{n,t}]$ is the velocity of target $n$ at time $t$.
The state transition matrices corresponding to the CV and CT model are given by
\begin{equation}\label{eq2}
\bm{F}_{n}^{\text{CV}}=\bm{\mathrm{I}}_2 \otimes
\begin{bmatrix}
1 & T_s \\
0 & 1 \\
\end{bmatrix},
\end{equation}
\begin{equation}\label{eqAddfct}
\bm{F}_{n}^{\text{CT}}=\begin{bmatrix}
    1 & \dfrac{\sin\left(\omega T_s\right)}{\omega} & 0 & \dfrac{\cos\left(\omega T_s\right)-1}{\omega} \\
    0 & \cos\left(\omega T_s\right) & 0 & \sin\left(\omega T_s\right) \\
    0 & \dfrac{1-\cos\left(\omega T_s\right)}{\omega} & 1 & \dfrac{\sin\left(\omega T_s\right)}{\omega} \\
    0 & -\sin\left(\omega T_s\right) & 0 & \cos\left(\omega T_s\right) \\
\end{bmatrix},
\end{equation}
where $T_s=1$ denotes the tracking time interval, $\omega$ denotes the turn rate, and $\otimes$ represents the Kronecker product.

The process noise variance matrix
\begin{equation}\label{eqAddQ}
\bm{\mathrm{Q}}_{n}^{m}=q_n^m \bm{\mathrm{I}}_2 \otimes
\begin{bmatrix}
    T_s^3/3 & T_s^2/2  \\
    T_s^2/2 & T_s \\
\end{bmatrix},
\end{equation}
where $q_n^m$ denotes the amplitude of the process noise. $\bm{\mathrm{I}}_2$ denotes the $2\times 2$ identity matrix.

The measurement matrix $\bm{\mathrm{H}}_n$ is
\begin{equation}\label{eq4}
\bm{\mathrm{H}}_n=\begin{bmatrix}
1 & 0 & 0 & 0\\
0 & 0 & 1 & 0\\
\end{bmatrix}.
\end{equation}

We assume that $\omega=3^{\circ}$, $q_n^{\text{CV}}=1$, $\bm{\mathrm{R}}_n=2*\bm{\mathrm{I}}_2$, and measurement cost $h_n=0$, $n=1,\ldots,N$.
The initial state $\mathbf{P}_{n,0}$ is generated by $\bm{\mathrm{R}}_0'\bm{\mathrm{R}}_0$, where each element of $\bm{\mathrm{R}}_0$ follows the uniform distribution $\mathrm{U}(0,1)$, ensuring that $\mathbf{P}_{n,0}\in\mathbb{R}^{4\times 4}_+$.
Then, we define two scenarios with the reckless type and the cautious type of targets with $q_n^{\text{CT}}=\{2,3,\ldots,9\}$, respectively, and another scenario with two types of targets with $q_n^{\text{CT}}=\{2,3,4,5,2,3,4,5\}$, where the number of each type of targets is 4, respectively.
In the first two scenarios, the weight is assumed as $d_n=5$, $n=1$, and $d_n=1$, $n\neq1$.
Otherwise, $d_n=5$ for reckless targets and $d_n=1$ for cautious targets in the third scenario.

The results with $N_{mc}=100$ Monte Carlo simulations are shown in Table~\ref{tab:q4}.
As the value of $K$ varies from 1 to 3,
While the number of targets $N$ remains constant at 8, the number of targets tracked will increase.
Consequently, the tracking costs obtained by all policies naturally decrease, and the difference between the MP index policy and other greedy policies decreases.
The MP index policy based on the trace of TEC matrices always obtains lower tracking errors and outperforms the TEC index policy and myopic policy.
\begin{table}[!t]
\begin{center}
\caption{{Results with different numbers of radars for different type targets with multi-dimensional states}}
\label{tab:q4}
\begin{tabular}{ c | c | c | c | c }
\hline
Type & Policies & $K=1$ & $K=2$ & $K=3$ \\
\hline
\multirow{3}{*}{Reckless}
& MP index & 4364.18 & 1142.87 & 610.10 \\
& myopic & 4480.28 & 1153.19 & 613.15 \\
& TEC & 4436.27 & 1153.43 & 633.80  \\
\hline
\multirow{3}{*}{Cautious}
& MP index & 3468.00 & 902.06 & 492.69 \\
& myopic & 3584.63 & 931.63 & 500.40 \\
& TEC & 3534.40 & 917.27 & 504.08  \\
\hline
\multirow{3}{*}{\makecell{Reckless and \\ Cautious}}
& MP index & 6777.79 & 1824.24 & 990.12  \\
& myopic & 7014.92 & 1895.25 & 1022.45  \\
& TEC & 6879.65 & 1860.21 & 1040.56  \\
\hline
\end{tabular}
\end{center}
\end{table}

\begin{figure}[!t]
\centering
\includegraphics[width=2.2in]{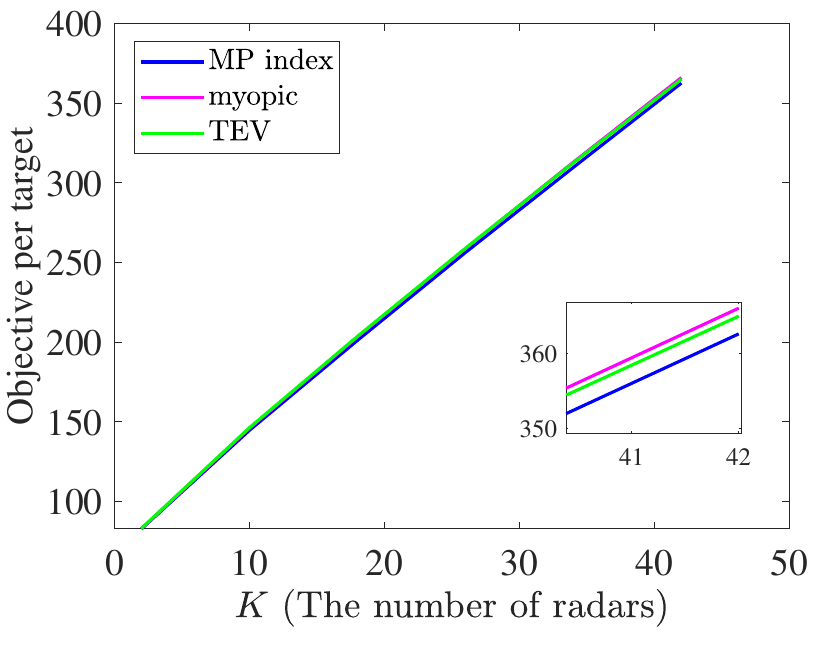}
\caption{Performance for two types of targets with $q_{n}^{\text{CT}}=\{2,\ldots, N/2+1, 2,\ldots, N/2+1\}$ and $d_n\equiv1$.}
\label{fig6}
\end{figure}

\begin{figure}[!t]
\centering
\includegraphics[width=2.2in]{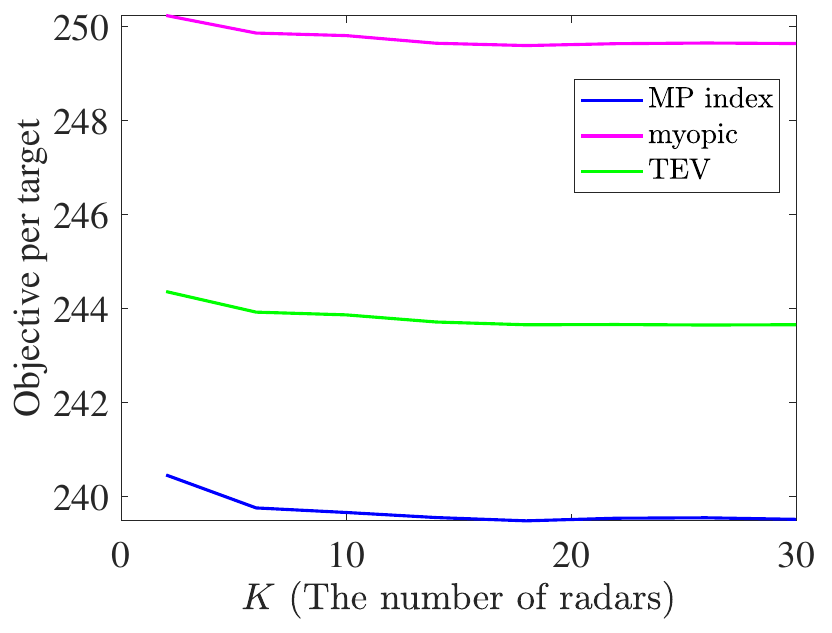}
\caption{Performance for two types of targets with $q_{n}^{\text{CT}}\equiv4$ and $d_n=5$ only for reckless targets.}
\label{fig7}
\end{figure}

Here, we also analyze the performance of policies changing over the number of targets and radars.
The initial state $\mathbf{P}_{n,0}$ is generated by $\bm{\mathrm{R}}_0'\bm{\mathrm{R}}_0$, where each element of $\bm{\mathrm{R}}_0$ follows the uniform distribution $\mathrm{U}(0,1)$, ensuring that $\mathbf{P}_{n,0}\in\mathbb{S}^{4}_{++}$.
The two scenarios are considered, where there are $50\%$ reckless targets and $50\%$ cautious targets, respectively.
In the first scenario, $q_{n}^{\text{CT}}=\{2,\ldots, N/2+1\}$ in each type of targets, and $d_n=1$ for all of the targets.
In the second scenario, the process noise variance $q_{n}^{\text{CT}}=4$ for all of the targets, but $d_n=5$ particularly for reckless targets.
The performance results under $N_{mc}=100$ Monte Carlo simulations are shown in Fig.~\ref{fig6} and Fig.~\ref{fig7}.
As the number of radars and targets increases, the MP index policy obtains better tracking performance in all of the cases.

These simulation results of targets with multi-dimensional TEC states show that the MP index policy also can efficiently solve the beam scheduling problem. However, the indexability validation and the lower bound need further calculations and verifications.

Above all, the adapted MP index policy improves the non-myopic beam scheduling performance and outperforms other greedy policies. Numerical simulation results with different parameters of the problem model indicate the superiority of this policy.

\section{Conclusion}\label{sec:conclusion}
Considering the beam scheduling problem for multiple smart target tracking, we have formulated each target as an MDP, where the state transition is established based on the Kalman filter and the dynamics model probability parameters.
Then the trade-off, which is between more observation of smart targets for better tracking performance and the maneuvering reaction of smart targets to elude themselves, was solved by the RMABP problem model based on the Whittle index policy.
Through numerical simulation results for both one-dimensional and multi-dimensional TEC states, we have demonstrated the better performance of the MP index policy by comparing it with the baseline greedy policies.
For future research, we plan to prompt the resource scheduling for smart targets tacking with multi-dimensional states in the RMABP model and prove the validity of the Whittle index policy by theoretical analysis.

\bibliographystyle{IEEEtran}
\bibliography{IEEEabrv,reference}

\begin{thebibliography}{10}
\providecommand{\url}[1]{#1}
\csname url@samestyle\endcsname
\providecommand{\newblock}{\relax}
\providecommand{\bibinfo}[2]{#2}
\providecommand{\BIBentrySTDinterwordspacing}{\spaceskip=0pt\relax}
\providecommand{\BIBentryALTinterwordstretchfactor}{4}
\providecommand{\BIBentryALTinterwordspacing}{\spaceskip=\fontdimen2\font plus
\BIBentryALTinterwordstretchfactor\fontdimen3\font minus
  \fontdimen4\font\relax}
\providecommand{\BIBforeignlanguage}[2]{{%
\expandafter\ifx\csname l@#1\endcsname\relax
\typeout{** WARNING: IEEEtran.bst: No hyphenation pattern has been}%
\typeout{** loaded for the language `#1'. Using the pattern for}%
\typeout{** the default language instead.}%
\else
\language=\csname l@#1\endcsname
\fi
#2}}
\providecommand{\BIBdecl}{\relax}
\BIBdecl

\bibitem{haimovich2007mimo}
A.~M. Haimovich, R.~S. Blum, and L.~J. Cimini, ``{MIMO} radar with widely
  separated antennas,'' \emph{IEEE Signal Processing Magazine}, vol.~25, no.~1,
  pp. 116--129, 2007.

\bibitem{zhang2022joint}
W.~Zhang, C.~Shi, J.~Zhou, and R.~Lv, ``Joint aperture and transmit resource
  allocation strategy for multi-target localization in phased array radar
  network,'' \emph{IEEE Transactions on Aerospace and Electronic Systems},
  vol.~59, no.~2, pp. 1551--1565, 2023.

\bibitem{yan2022radar}
J.~Yan, H.~Jiao, W.~Pu, C.~Shi, J.~Dai, and H.~Liu, ``Radar sensor network
  resource allocation for fused target tracking: a brief review,''
  \emph{Information Fusion}, vol.~86, pp. 104--115, 2022.

\bibitem{kreucher2006adaptive}
C.~Kreucher, D.~Blatt, A.~Hero, and K.~Kastella, ``Adaptive multi-modality
  sensor scheduling for detection and tracking of smart targets,''
  \emph{Digital Signal Processing}, vol.~16, no.~5, pp. 546--567, 2006.

\bibitem{savage2009sensor}
C.~O. Savage and B.~F. La~Scala, ``Sensor management for tracking smart
  targets,'' \emph{Digital Signal Processing}, vol.~19, no.~6, pp. 968--977,
  2009.

\bibitem{zhang2015non}
Z.~Zhang and G.~Shan, ``Non-myopic sensor scheduling to track multiple reactive
  targets,'' \emph{IET Signal Processing}, vol.~9, no.~1, pp. 37--47, 2015.

\bibitem{hero2007foundations}
A.~O. Hero, D.~Casta{\~n}{\'o}n, D.~Cochran, and K.~Kastella, \emph{Foundations
  and applications of sensor management}.\hskip 1em plus 0.5em minus
  0.4em\relax Springer Science \& Business Media, 2007.

\bibitem{taner2012scheduling}
M.~R. Taner, O.~E. Karasan, and E.~Yavuzturk, ``Scheduling beams with different
  priorities on a military surveillance radar,'' \emph{IEEE Transactions on
  Aerospace and Electronic Systems}, vol.~48, no.~2, pp. 1725--1739, 2012.

\bibitem{godrich2011sensor}
H.~Godrich, A.~P. Petropulu, and H.~V. Poor, ``Sensor selection in distributed
  multiple-radar architectures for localization: A knapsack problem
  formulation,'' \emph{IEEE Transactions on Signal Processing}, vol.~60, no.~1,
  pp. 247--260, 2011.

\bibitem{zhang2022dynamic}
G.~Zhang, J.~Xie, H.~Zhang, Z.~Li, and C.~Qi, ``Dynamic antenna selection for
  colocated {MIMO} radar,'' \emph{Remote Sensing}, vol.~14, no.~12, p. 2912,
  2022.

\bibitem{zhang2020antenna}
H.~Zhang, J.~Shi, Q.~Zhang, B.~Zong, and J.~Xie, ``Antenna selection for target
  tracking in collocated {MIMO} radar,'' \emph{IEEE Transactions on Aerospace
  and Electronic Systems}, vol.~57, no.~1, pp. 423--436, 2020.

\bibitem{dai2022adaptive}
J.~Dai, J.~Yan, W.~Pu, H.~Liu, and M.~S. Greco, ``Adaptive channel assignment
  for maneuvering target tracking in multistatic passive radar,'' \emph{IEEE
  Transactions on Aerospace and Electronic Systems}, vol.~59, no.~3, pp.
  2780--2793, 2023.

\bibitem{yi2020resource}
W.~Yi, Y.~Yuan, R.~Hoseinnezhad, and L.~Kong, ``Resource scheduling for
  distributed multi-target tracking in netted colocated {MIMO} radar systems,''
  \emph{IEEE Transactions on Signal Processing}, vol.~68, pp. 1602--1617, 2020.

\bibitem{zhang2020joint}
H.~Zhang, W.~Liu, J.~Xie, Z.~Zhang, and W.~Lu, ``Joint subarray selection and
  power allocation for cognitive target tracking in large-scale {MIMO} radar
  networks,'' \emph{IEEE Systems Journal}, vol.~14, no.~2, pp. 2569--2580,
  2020.

\bibitem{xie2017joint}
M.~Xie, W.~Yi, T.~Kirubarajan, and L.~Kong, ``Joint node selection and power
  allocation strategy for multitarget tracking in decentralized radar
  networks,'' \emph{IEEE Transactions on Signal Processing}, vol.~66, no.~3,
  pp. 729--743, 2017.

\bibitem{shi2021joint}
C.~Shi, Y.~Wang, S.~Salous, J.~Zhou, and J.~Yan, ``Joint transmit resource
  management and waveform selection strategy for target tracking in distributed
  phased array radar network,'' \emph{IEEE Transactions on Aerospace and
  Electronic Systems}, vol.~58, no.~4, pp. 2762--2778, 2021.

\bibitem{yan2020optimal}
J.~Yan, W.~Pu, S.~Zhou, H.~Liu, and M.~S. Greco, ``Optimal resource allocation
  for asynchronous multiple targets tracking in heterogeneous radar networks,''
  \emph{IEEE Transactions on Signal Processing}, vol.~68, pp. 4055--4068, 2020.

\bibitem{yan2020collaborative}
J.~Yan, W.~Pu, S.~Zhou, H.~Liu, and Z.~Bao, ``Collaborative detection and power
  allocation framework for target tracking in multiple radar system,''
  \emph{Information Fusion}, vol.~55, pp. 173--183, 2020.

\bibitem{pang2019sensor}
C.~Pang and G.~Shan, ``Sensor scheduling based on risk for target tracking,''
  \emph{IEEE Sensors Journal}, vol.~19, no.~18, pp. 8224--8232, 2019.

\bibitem{krishn16}
V.~Krishnamurthy, \emph{Partially Observed Markov Decision Processes: From
  Filtering to Controlled Sensing}.\hskip 1em plus 0.5em minus 0.4em\relax
  Cambridge, UK: Cambridge University Press, 2016.

\bibitem{kreucher2004efficient}
C.~Kreucher, A.~O. Hero, K.~Kastella, and D.~Chang, ``Efficient methods of
  non-myopic sensor management for multitarget tracking,'' in \emph{2004 43rd
  IEEE Conference on Decision and Control (CDC)(IEEE Cat. No. 04CH37601)},
  vol.~1.\hskip 1em plus 0.5em minus 0.4em\relax IEEE, 2004, pp. 722--727.

\bibitem{kreucher2006monte}
C.~M. Kreucher and A.~O. Hero, ``{Monte Carlo} methods for sensor management in
  target tracking,'' in \emph{2006 IEEE Nonlinear Statistical Signal Processing
  Workshop}.\hskip 1em plus 0.5em minus 0.4em\relax IEEE, 2006, pp. 232--237.

\bibitem{gongguo2019non}
X.~Gongguo, S.~Ganlin, and D.~Xiusheng, ``Non-myopic scheduling method of
  mobile sensors for manoeuvring target tracking,'' \emph{IET Radar, Sonar \&
  Navigation}, vol.~13, no.~11, pp. 1899--1908, 2019.

\bibitem{shan2020non}
G.~Shan, G.~Xu, and C.~Qiao, ``A non-myopic scheduling method of radar sensors
  for maneuvering target tracking and radiation control,'' \emph{Defence
  Technology}, vol.~16, no.~1, pp. 242--250, 2020.

\bibitem{dong2021risk}
Q.~Dong and C.~Pang, ``Risk-based non-myopic sensor scheduling in target threat
  level assessment,'' \emph{IEEE Access}, vol.~9, pp. 76\,379--76\,394, 2021.

\bibitem{shan2017non}
G.~Shan and Z.~Zhang, ``Non-myopic sensor scheduling for low radiation risk
  tracking using mixed {POMDP},'' \emph{Transactions of the Institute of
  Measurement and Control}, vol.~39, no.~2, pp. 230--243, 2017.

\bibitem{howard2004optimal}
S.~Howard, S.~Suvorova, and B.~Moran, ``Optimal policy for scheduling of
  {Gauss--Markov} systems,'' in \emph{Proceedings of the Seventh International
  Conference on Information Fusion}.\hskip 1em plus 0.5em minus 0.4em\relax
  Citeseer, 2004, pp. 888--892.

\bibitem{whittle1988restless}
P.~Whittle, ``Restless bandits: Activity allocation in a changing world,''
  \emph{Journal of Applied Probability}, vol.~25, no.~A, pp. 287--298, 1988.

\bibitem{gittins1979bandit}
J.~C. Gittins, ``Bandit processes and dynamic allocation indices,''
  \emph{Journal of the Royal Statistical Society: Series B (Methodological)},
  vol.~41, no.~2, pp. 148--164, 1979.

\bibitem{krishnamurthy2001hidden}
V.~Krishnamurthy and R.~J. Evans, ``Hidden {Markov} model multiarm bandits: A
  methodology for beam scheduling in multitarget tracking,'' \emph{IEEE
  Transactions on Signal Processing}, vol.~49, no.~12, pp. 2893--2908, 2001.

\bibitem{gittins2011multi}
J.~Gittins, K.~Glazebrook, and R.~Weber, \emph{Multi-armed bandit allocation
  indices}.\hskip 1em plus 0.5em minus 0.4em\relax John Wiley \& Sons, 2011.

\bibitem{papTsik99}
C.~H. Papadimitriou and J.~N. Tsitsiklis, ``The complexity of optimal queuing
  network control,'' \emph{Mathematics of Operations Research}, vol.~24, no.~2,
  pp. 293--305, 1999.

\bibitem{nmmath23}
J.~Ni{\~n}o-Mora, ``Markovian restless bandits and index policies: {A}
  review,'' \emph{Mathematics}, vol.~11, no.~7, p. 1639, 2023.

\bibitem{weber1990index}
R.~R. Weber and G.~Weiss, ``On an index policy for restless bandits,''
  \emph{Journal of applied probability}, vol.~27, no.~3, pp. 637--648, 1990.

\bibitem{la2006optimal}
B.~F. La~Scala and B.~Moran, ``Optimal target tracking with restless bandits,''
  \emph{Digital Signal Processing}, vol.~16, no.~5, pp. 479--487, 2006.

\bibitem{dance2015kalman}
C.~R. Dance and T.~Silander, ``When are {K}alman-filter restless bandits
  indexable?'' in \emph{Proceedings of the 28th Conference on Neural
  Information Processing Systems (NIPS), Montreal, Canada}.\hskip 1em plus
  0.5em minus 0.4em\relax Cambridge, MA: MIT Press, 2015, pp. 1711--1719.

\bibitem{liu2010indexability}
K.~Liu and Q.~Zhao, ``Indexability of restless bandit problems and optimality
  of whittle index for dynamic multichannel access,'' \emph{IEEE Transactions
  on Information Theory}, vol.~56, no.~11, pp. 5547--5567, 2010.

\bibitem{nino2011sensor}
J.~Ni{\~n}o-Mora and S.~S. Villar, ``Sensor scheduling for hunting elusive
  hiding targets via {Whittle's} restless bandit index policy,'' in
  \emph{International Conference on NETwork Games, Control and Optimization
  (NetGCooP 2011)}.\hskip 1em plus 0.5em minus 0.4em\relax IEEE, 2011, pp.
  1--8.

\bibitem{wang2019whittle}
J.~Wang, X.~Ren, Y.~Mo, and L.~Shi, ``Whittle index policy for dynamic
  multichannel allocation in remote state estimation,'' \emph{IEEE Transactions
  on Automatic Control}, vol.~65, no.~2, pp. 591--603, 2019.

\bibitem{nino2001restless}
J.~Ni{\~n}o-Mora, ``Restless bandits, partial conservation laws and
  indexability,'' \emph{Advances in Applied Probability}, vol.~33, no.~1, pp.
  76--98, 2001.

\bibitem{nmmp02}
------, ``Dynamic allocation indices for restless projects and queueing
  admission control: {A} polyhedral approach,'' \emph{Mathematical
  programming}, vol.~93, no.~3, pp. 361--413, 2002.

\bibitem{nmmor06}
------, ``Restless bandit marginal productivity indices, diminishing returns
  and optimal control of make-to-order/make-to-stock {M}/{G}/1 queues,''
  \emph{Mathematics of Operations Research}, vol.~31, no.~1, pp. 50--84, 2006.

\bibitem{nino2020verification}
------, ``A verification theorem for threshold-indexability of real-state
  discounted restless bandits,'' \emph{Mathematics of Operations Research},
  vol.~45, no.~2, pp. 465--496, 2020.

\bibitem{nmconsLawsEORMS}
------, ``Conservation laws and related applications,'' in \emph{Wiley
  Encyclopedia of Operations Research and Management Science}.\hskip 1em plus
  0.5em minus 0.4em\relax New York, NY, USA: Wiley Online Library, 2011.

\bibitem{nino2008index}
------, ``An index policy for dynamic fading-channel allocation to
  heterogeneous mobile users with partial observations,'' in \emph{Proceedings
  of the 4th EuroNGI Conference on Next Generation Internet Networks (NGI),
  Krakow, Poland}.\hskip 1em plus 0.5em minus 0.4em\relax New York, NY, USA:
  IEEE, 2008, pp. 231--238.

\bibitem{nino2009multitarget}
J.~Ni{\~n}o-Mora and S.~S. Villar, ``Multitarget tracking via restless bandit
  marginal productivity indices and {K}alman filter in discrete time,'' in
  \emph{Proceedings of the Joint 48th IEEE Conference on Decision and Control
  (CDC) / 28th Chinese Control Conference (CCC), Shanghai, China}.\hskip 1em
  plus 0.5em minus 0.4em\relax New York, NY, USA: IEEE, 2009, pp. 2905--2910.

\bibitem{nino2016whittle}
J.~Ni{\~n}o-Mora, ``Whittle’s index policy for multi-target tracking with
  jamming and nondetections,'' in \emph{Proceedings of the 23rd International
  Conference on Analytical and Stochastic Modelling Techniques and Applications
  (ASMTA), Cardiff, Wales}.\hskip 1em plus 0.5em minus 0.4em\relax Cham,
  Switzerland: Springer, 2016, pp. 210--222.

\bibitem{nino2015verification}
------, ``A verification theorem for indexability of discrete time real state
  discounted restless bandits,'' \emph{{a}rXiv:1512.04403}, 2015.

\bibitem{dance2019optimal}
C.~R. Dance and T.~Silander, ``Optimal policies for observing time series and
  related restless bandit problems,'' \emph{The Journal of Machine Learning
  Research}, vol.~20, no.~1, pp. 1218--1310, 2019.

\bibitem{yang2020restless}
F.~Yang and X.~Luo, ``A restless {MAB}-based index policy for {UL} pilot
  allocation in massive {MIMO} over {Gauss--Markov} fading channels,''
  \emph{IEEE Transactions on Vehicular Technology}, vol.~69, no.~3, pp.
  3034--3047, 2020.

\bibitem{visina2018multiple}
R.~Visina, Y.~Bar-Shalom, and P.~Willett, ``Multiple-model estimators for
  tracking sharply maneuvering ground targets,'' \emph{IEEE Transactions on
  Aerospace and Electronic Systems}, vol.~54, no.~3, pp. 1404--1414, 2018.

\bibitem{blair2023mse}
W.~D. Blair and Y.~Bar-Shalom, ``{MSE} design of nearly constant velocity
  {Kalman} filters for tracking targets with deterministic maneuvers,''
  \emph{IEEE Transactions on Aerospace and Electronic Systems}, vol.~59, no.~4,
  pp. 4180--4191, 2023.

\bibitem{brown2020index}
D.~B. Brown and J.~E. Smith, ``Index policies and performance bounds for
  dynamic selection problems,'' \emph{Management Science}, vol.~66, no.~7, pp.
  3029--3050, 2020.

\end{thebibliography}


\begin{IEEEbiographynophoto}{Yuhang Hao}
received the B.Sc. degree and the M.Sc. degree in control theory and control engineering from Northwestern Polytechnical University, Xi’an, China, in 2016 and 2019 respectively. He is currently pursuing the Ph.D. degree with the Key Laboratory of information Fusion Technology, Ministry of Education, Northwestern Polytechnical University, Xi’an, China. His research interests include resource scheduling, multi-armed bandits, target tracking, and information fusion.
\end{IEEEbiographynophoto}
\begin{IEEEbiographynophoto}{Zengfu Wang}
received the B.Sc. degree in applied mathematics, the M.Sc. degree in control theory and control engineering, the Ph.D. degree in control science and engineering from Northwestern Polytechnical University, Xi’an, China, in 2005, 2008, and 2013 respectively. From 2014 to 2017, he was a Lecturer with Northwestern Polytechnical University, where he is currently an Associate Professor. From 2014 to 2015, he was a Postdoctoral Research Fellow with the Department of Electronic Engineering, City University of Hong Kong, Hong Kong, China. Since December 2019, he has been a Visiting Researcher with the Faculty of Electrical Engineering, Mathematics and Computer Science, Delft University of Technology, Delft, The Netherlands. His research interests include path planning, discrete optimization, and information fusion.
\end{IEEEbiographynophoto}
\begin{IEEEbiographynophoto}{Jos\'{e} Ni\~{n}o-Mora}
earned a Lic. (B.Sc./M.Sc.)  degree in Mathematical Sciences from Complutense University in Madrid, Spain, and a Ph.D. degree in Operations Research from Massachusetts Institute of Technology (MIT) on a Fulbright fellowship, in 1989 and 1995, respectively.
After postdoc stints at MIT and at the Catholic University of Louvain at Louvain-la-Neuve, Belgium (as a Marie Curie fellow), he was a Visiting Professor at the
Economics \& Business Department of Pompeu Fabra University in Barcelona, and then joined in 2003 the Statistics Department at Carlos III University of Madrid, Spain, where he is Full Professor of Operations Research \& Statistics.
He was the recipient of the 1st Award for Best Methodological Contribution in Operations Research in 2020 by the Spanish Statistics and Operations Research Society and the BBVA Foundation.
His research interests include dynamic resource allocation in stochastic systems, Markov decision models, restless bandit problems, and index policies.
\end{IEEEbiographynophoto}
\begin{IEEEbiographynophoto}{Jing Fu} received the B.Eng. degree in computer science from Shanghai Jiao Tong University, Shanghai, China, in 2011, and the Ph.D. degree in electronic engineering from the City University of Hong Kong in 2016. She has been with the School of Mathematics and Statistics, the University of Melbourne as a Post-Doctoral Research Associate from 2016 to 2019. She has been a lecturer in the discipline of Electronic \& Telecommunications Engineering, RMIT University, since 2020. Her research interests now include energy-efficient networking/scheduling, resource allocation in large-scale networks, semi-Markov/Markov decision processes, restless multi-armed bandit problems, and stochastic optimization.
\end{IEEEbiographynophoto}
\begin{IEEEbiographynophoto}{Min Yang} received the B.Sc. degree from Shandong Agricultural University in 2015, Taian, China. She received the M.Sc. degree in Control science and engineering from University of Science and Technology Liaoning, Liaoning, China, in 2019. She is currently pursuing the Ph.D.degree with the Key Laboratory of information Fusion Technology, Ministry of Education, Northwestern Polytechnical University, Xi'an, China. Her research interests include multi-armed bandits, resource allocation, and target tracking.
\end{IEEEbiographynophoto}
\begin{IEEEbiographynophoto}{Quan Pan}
received the B.Sc. degree from Huazhong Institute of Technology in 1991, and the M.Sc. and Ph.D. degrees from Northwestern Polytechnical University in 1991 and 1997, respectively.
Since 1997, he has been a professor with the School of Automation, Northwestern Polytechnical University.
His research interests include information fusion, hybrid system estimation theory, multi-scale estimation theory, target tracking and image processing.
He is a Member of the International Society of Information Fusion, a Board Member of the Chinese Association of Automation, and a Member of the Chinese Association of Aeronautics and Astronautics.
He was the recipient of the 6th Chinese National Youth Award for Outstanding Contribution to Science and Technology in 1998 and the Chinese National New Century Excellent Professional Talent in 2000.
\end{IEEEbiographynophoto}

\vfill

\end{document}